# Unconventional mixed state in the nematic superconductor LiFeAs

G. Lamura,[1, *] T. Winyard,[2, 3, †] P. Gentile,[4, ‡] M. Speight,[5] F. Anger,[6] B. Büchner,[6] S. Wurmehl,[6] and T. Shiroka[7, 8, §]

[1]*CNR-SPIN, I-16152 Genova, Italy*

[2]*Division of Mathematics, University of Dundee, Dundee DD1 4HN, United Kingdom*

[3]*Maxwell Institute, University of Edinburgh, Edinburgh EH9 3FD, United Kingdom*

[4]*CNR-SPIN, c/o Università degli Studi di Salerno, I-84084 Fisciano (SA), Italy*

[5]*School of Mathematics, University of Leeds, Leeds LS2 9JT , United Kingdom*

[6]*Leibniz-Institute for Solid State and Materials Research, IFW-Dresden, 01069 Dresden, Germany*

[7]*PSI Center for Neutron and Muon Sciences CNM, CH-5232 Villigen PSI, Switzerland*

[8]*Laboratorium für Festkörperphysik, ETH Zürich, CH-8093 Zürich, Switzerland*

In the mixed state of type-II bulk superconductors, the magnetic field penetrates in the form of vortices enclosing one magnetic flux quantum: this is the conventional Abrikosov vortex lattice. Here, by using transverse muon-spin spectroscopy, we demonstrate the presence of an unconventional vortex lattice in LiFeAs single crystals. We also show evidence that the new mixed phase consists of stripes of "coreless" vortices, which are bound states of two spatially separated half-quantum vortices.

## I. INTRODUCTION

In a type II superconductor, the mixed state, also called the Abrikosov vortex state [1], is the regime between two critical magnetic fields, the lower $H_{c1}$ and the upper $H_{c2}$, where the magnetic flux partially penetrates the material in quantized vortices. Such magnetic excitations are characterized by a normal-state core of radius $\xi$, surrounded by screening supercurrents over a distance $\lambda_{\rm L}$, each carrying one quantum of magnetic flux $\phi_0 = h/2e$, being $e$ and $h$ the electron charge and Planck's constant, respectively. They arrange themselves into the so-called Abrikosov vortex lattice (AVL). This picture applies to superconductors with even-parity pairing, where Cooper pairs form a spin-singlet state and the order parameter (OP) is a scalar function [2].

In contrast, for odd-parity (triplet) pairing, the gap function is generally expressed as

$$\Delta(\mathbf{k}) = i \left( \mathbf{d}(\mathbf{k}) \cdot \hat{\sigma} \right) \hat{\sigma}_y, \tag{1}$$

where $\mathbf{d}(\mathbf{k})$ encodes the spin configuration of the Cooper pairs [2]. This vectorial structure allows for a rich variety of superconducting states, including multiband, topological, and nematic phases depending on symmetry and interactions. Since the spin-triplet order parameter is characterized by multiple internal components, additional characteristic length scales and anisotropies [2, 3] emerge, which can qualitatively modify the mixed state and vortex interactions [4, 5].

In particular, while in conventional Abrikosov vortices a single complex order parameter winds by $2\pi$ around a singular core, in spin triplet superconductors the phase winding can be distributed among the different OP components. This gives rise to fractional vortices (FV), in which each component carries only part of the total winding, like in the case of half-quantum vortices (HQV), which carry half of the usual flux. The same internal degrees of freedom also enable coreless vortices (CLV). In such a case, instead of vanishing, the OP remains finite inside the core, while its internal structure continuously rotates in space. Such configurations are naturally stabilized when the degeneracy between components is lifted, for example, by nematic order, which selects a preferred orientation and allows spatial textures to replace the singular core [6, 7]. For an order parameter with $n$ components, the corresponding CLV behave like $n$ spatially separated $1/n$-quantum vortices, each of which contributes approximately $\Phi_0/n$ to the quantized local magnetic field. For a two-component system ($n = 2$), each coreless vortex consists of two bound HQV, as shown in Figure 1, where HQVs are graphically represented with their screening supercurrents and their independent OP amplitudes and phases. The screening supercurrent of each HQV overlaps with the core of the second HQV, and vice-versa. This results in a coreless vortex-bounded structure, where even if each individual OP vanishes, they never do so simultaneously. Such unconventional vortices are also named *skyrmion vortices*, because they carry a skyrmionic topological charge of 1 [8].

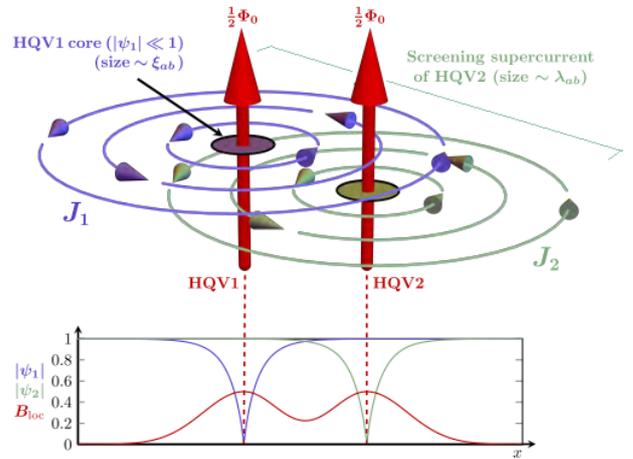

FIG. 1. **Coreless (skyrmion) vortex,** consisting of a bound state of two spatially separated HQVs with independent order parameters and phases. The cores of the two HQVs are spanned by the respective supercurrents $J_1$ and $J_2$: HQV1 is crossed by $J_2$ and visa-versa, thus lacking a core where both OPs are zero simultaneously. This can be seen in the bottom panel as a cross-section in the splitting direction of the two independent OPs and the resulting local field $B_{\rm loc}$ is shown. Such bound state still carries a single quantum of magnetic flux $\Phi_0$ as a whole.

Although non Abrikosov vortex lattices have been theoretically predicted, experimental evidence of them remains scarce and are limited to few cases. Using scanning SQUID magnetometry, (i) evidences of HQV have been found at the tricrystal point of $YBa_2Cu_3O_{7-\delta}$ grown on tricrystal substrates [9] and (ii) temperature-dependent FV have been detected in hole-overdoped $Ba_{1-x}K_xFe_2As_2$ ($x = 0.77$) on cleaved single crystals [10]. (iii) Observations of HQV have been reported in micrometer sized rings of the odd-parity superconductor $\beta$-$Bi_2Pd$ in the form of textured thin films via the Little–Parks experiment [11]. However authors attributed this evidence to the presence of grains and grain boundaries that could pin FV, as theoretically predicted [2]. (iv) Evidence of HQV have been found by cantilever magnetometry also in $Sr_2RuO_4$ single crystals containing a mesoscopic hole fabricated by focused ion-beam lithography [12]. The mesoscopic size of the sample significantly reduces the energy difference between integer and half-integer vortex states, making the latter energetically favorable [13–15]. The fact that FV and HQV have only been observed in mesoscopic systems or at crystal surfaces underscores the lack of any bulk evidence. Relevantly, CLV have been experimentally demonstrated in multicomponent superfluids such as superfluid $^3$He [16–18] and spinor Bose–Einstein condensates [19–22]. However, direct and unambiguous experimental evidence for CLV in superconducting materials is still lacking. Here, we show results compatible with their occurrence in the SC state of stoichiometric LiFeAs using angle-resolved muon-spin spectroscopy (μSR). We complement our experimental results with a theoretical description which is fully consistent with the CLV framework, and takes into account the specific features of LiFeAs SC state, in particular the spontaneous nematic order observed below $T_c \simeq 17$ K, via angle-resolved photoemission spectroscopy (ARPES) [23].

## II. ANGLE RESOLVED TF-μSR

We performed systematic angle-resolved transverse field muon spectroscopy (TF-μSR) measurements, by changing the relative orientation between the crystal axes and the magnetic field $B_{app}$ (see Fig. 2A (i)).

**In-plane TF-μSR**

A magnetic field of 14.5 mT was applied parallel to the $a$-axis ($B_{a0}$, Fig. 2A (ii)) and at 45° from it ($B_{a45}$, Fig. 2A (iii)) by using a field cooling (FC) protocol. Representative TF-μSR spectra collected in the normal- and superconducting states of LiFeAs are shown in Fig. 2B-(a,b). In the superconducting state (i.e., $T < T_c$), the development of a flux-line lattice (FLL) causes an inhomogeneous field distribution, which produces an additional damping in the TF-μSR spectra [24]. Figure 2B-(d) shows the Fourier transform (FT) of both low-temperature spectra. Here, the FT of the time-dependent polarization represents the probability distribution density of the local field $p(B)$. In general, in the superconducting state, the $p(B)$ distribution is material-dependent. When $p(B)$ is symmetric, one Gaussian distribution is sufficient to describe the TF-μSR spectra. In the present case, however, the superposition of two Gaussian distributions is needed to model the superconducting fraction, while another Gaussian is required to model the background [25]. The first and the second moment of such distribution represent the local field diamagnetic shift $B_\mu$ and the depolarization rate $\sigma$ [26]. No assumptions on the vortex lattice nature have been done at this stage. Figure 3a shows the temperature evolution of the depolarization rates $\sigma_{a0}$ and $\sigma_{a45}$. Although the low- and high-temperature values are nearly identical, the medium temperatures $\sigma(T)$ behavior is quite different for the two in-plane fields. This difference is even more pronounced in the field shift $\Delta B$, shown in Fig. 3b. Here, the simple diamagnetic saturation observed for $B_{a0}$ differs greatly from that observed for $B_{a45}$, for which a paramagnetic shift appears below 7 K. This behavior is inconsistent with a tetragonal crystal and *suggests a rotational symmetry breaking* (RSB) for rotations around the $c$-axis.

**Out-of-plane TF-μSR**

A magnetic field of 30 mT was applied parallel to the $c$-axis ($B_c$) using the FC protocol (Fig. 2A (iii)). Figures 2B-(c,e) show representative TF-μSR spectra collected in the normal- and superconducting states, as well as their FT. We note two interesting features: (i) a sharp peak at the applied-field value, representative of the muons implanted in the non-superconducting part of the sample (and/or parts of the sample holder). This gives the lower bound of the SC volume fraction $V_{SC} \simeq 89(1)$ % [25]. (ii) The superconducting contribution is represented by a split peak, diamagnetically shifted with respect to the applied field $B_c$. This totally unexpected *splitting* persists independently of the type of data analysis and evolves into a square-shaped peak when the FT apodization is increased. This key feature is particularly relevant because it indicates a nontrivial spectrum consisting of two near-lying frequencies, which we discuss in detail below. The time-dependent fitting of the measured polarization fails to capture this double feature (due to the limited muon life time) [27]. Thus, to extract the second moment $\langle \Delta B^2 \rangle$ of the internal field distribution, we considered a skewed Gaussian function, $skG(t)$ [28], for the superconducting component, since its FT consists of an asymmetric field distribution, which is best suited for the present case [25]. We show the temperature evolution of the resulting depolarization rate $\sigma_c$ and of the diamagnetic field shift $\Delta B_c$ in Figures 3-(a) and (b), respectively. In both cases, the low-$T$ limits are almost twice those measured in the *in*-plane field configurations. Here, $\sigma_c > \sigma_{a\theta}$ suggests a smaller in-plane magnetic penetration depth, hence, a stronger superconductivity. As for the shift, $|\Delta B_c| > |\Delta B_{a\theta}|$ simply reflects the applied magnetic field along the $c$-axis being twice that applied in the plane.

## III. DISCUSSION

The central result of our analysis is contained in Figure 2B-(e), and is represented by the peak splitting in the FT of the time-dependent polarization recorded at 2 K, when the magnetic field is oriented along the LiFeAs $c$-axis. We fitted the FT amplitude with a sum of Gaussian distributions, allowing us to precisely model the splitting: the green dotted lines represent the two peaks related to the superconducting contribution. We could follow their temperature evolution, as shown in Fig. 2B-(f). Interestingly, the local field splitting is initially absent at the lowest temperature (1.6 K), then it clearly appears at 2 K, to gradually vanish above 6 K.

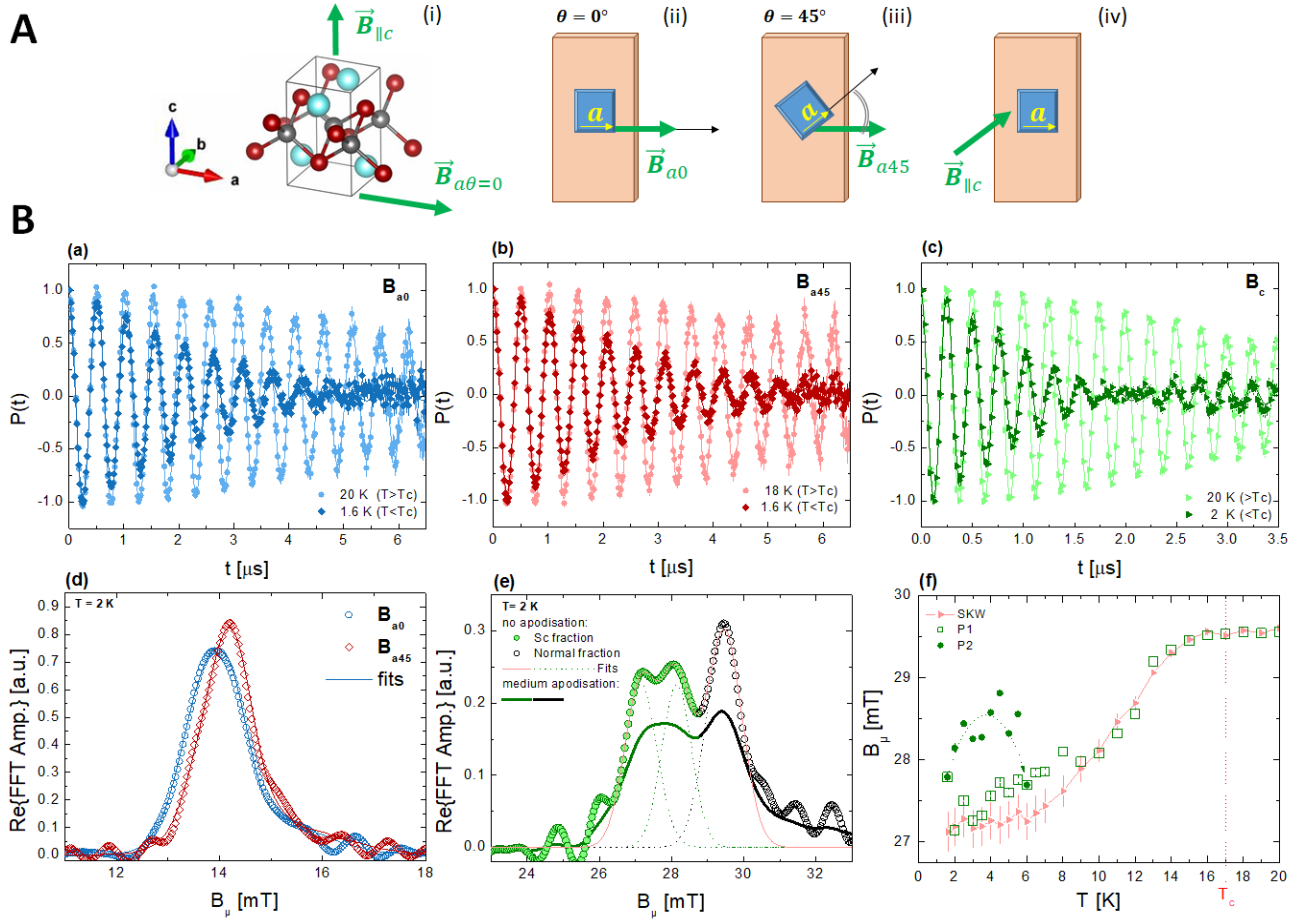

FIG. 2. **TF-µSR experiments.** Panel A: Crystal structure of LiFeAs (cyan, gray, and red spheres represent Li, Fe, and As atoms, respectively) and relative directions of the applied magnetic field(i). Experimental configuration: the sample is glued using vacuum grease to a thin copper foil on the µSR sample-holder. This fixes its $a$-axis direction at low temperature, while allowing it to be changed at 300 K. The magnetic field $B_{app}$ was applied parallel to the $a$-axis (ii), at 45° from it (iii), and parallel to the $c$-axis (iv).
Panel B: TF-µSR polarization above and below $T_c$ in a magnetic field applied parallel to the $a$-axis ($B_{a0}$) (a), at 45 degrees from it ($B_{45}$) (b), and parallel to the $c$-axis ($B_c$) (c). The fast Fourier transforms (FFTs) of the time-domain spectra in the SC phase (at 2 K) are shown in panels (d) and (e) with weak and no apodization, respectively. In particular, in panel (e), the green and black circles represent the SC and N fractions, with the dotted lines representing the superconducting- and normal-state fit components (the red line represents the full fit). A clear splitting of the superconducting peak is visible. The continuous green and black lines represents the FFT of the same data set with a medium level apodization. The flat-top shape of the SC peak indicates the contribution of two near yet distinct frequencies. Panel (f) shows the temperature evolution of the peak splitting shown in (e). Note the disappearance of peak splitting above 6 K. Full squares represent the average local field obtained when using a SKW model. See text for details.

Due to the demonstrated nematic character of the SC OP of LiFeAs, the observed peak splitting in the local field distribution $p(B)$ can be explained as a manifestation of a skyrmionic vortex lattice [6].

Indeed, differently from the conventional mixed phase, where the vortex lattice depends only on the intervortex distance, skyrmionic vortices are expected to be arranged in stripes, described by two length scales: the separation between the stripes and the separation of skyrmions within the same stripe. It has been theoretically demonstrated that each of these two length scales gives rise to a distinct peak in the field probability distribution [6]. For strong applied fields, the fractional vortices merge to form a standard vortex lattice, thereby losing the second peak in $p(B)$. A crucial condition for the emergence of a skyrmionic vortex lattice, and consequently for the appearance of a double-peak structure in the local field distribution, is that the underlying SC state is nematic and belongs to a two-dimensional representation, with finite $d_x$ and $d_y$ components [6]. The nematic order lifts the degeneracy between these components, selecting a preferred in-plane orientation and enabling nontrivial vortex textures. In the following, we provide a theoretical analysis of the possible SC states of LiFeAs that satisfy such conditions, and then by exploiting a Giunburg-Landau model we show that the temperature evolution found in our experiments for the peak splitting of $p(B)$ mirrors the predicted field evolution, as the CLV lattice melts into an AVL when temperature is increased.

*Superconducting order parameter*

Although LiFeAs does not exhibit nematic fluctuations [29], magnetic transitions [30], or structural transitions as a function of temperature or applied pressure (up to 17 GPa) [31, 32], the *in*-plane ARPES data in the SC phase exhibit a twofold ($C_2$) rotational symmetry instead of the fourfold ($C_4$) symmetry expected in a tetragonal system [23]. Our TF-µSR results further confirm the nematic SC ground state of LiFeAs when fields are applied in the *ab*-plane. To fully characterize the SC state of LiFeAs, its low-energy electronic structure has to be considered, which is dominated by Fe 3*d* states and comprises multiple hole- and electron-like bands that cross the Fermi level. Two hole pockets, the $\alpha$ and $\beta$ bands, are present at the Brillouin-zone center ($\Gamma$ point). The inner $\alpha$ band primarily consists of Fe $d_{xz}/d_{yz}$ orbitals. The outer $\beta$ band produces a larger hole pocket formed by a mixture of $d_{xz}/d_{yz}$ and $d_{xy}$ orbitals. Both bands hybridize weakly with the As $p_z$ orbital, which introduces a small $k_z$ dispersion. At the Brillouin-zone corners, two electron-like bands ($\gamma$ and $\delta$) give rise to two electron pockets with mainly Fe $d_{xz}/d_{yz}$ orbitals and some admixture of $d_{xy}$ orbitals.

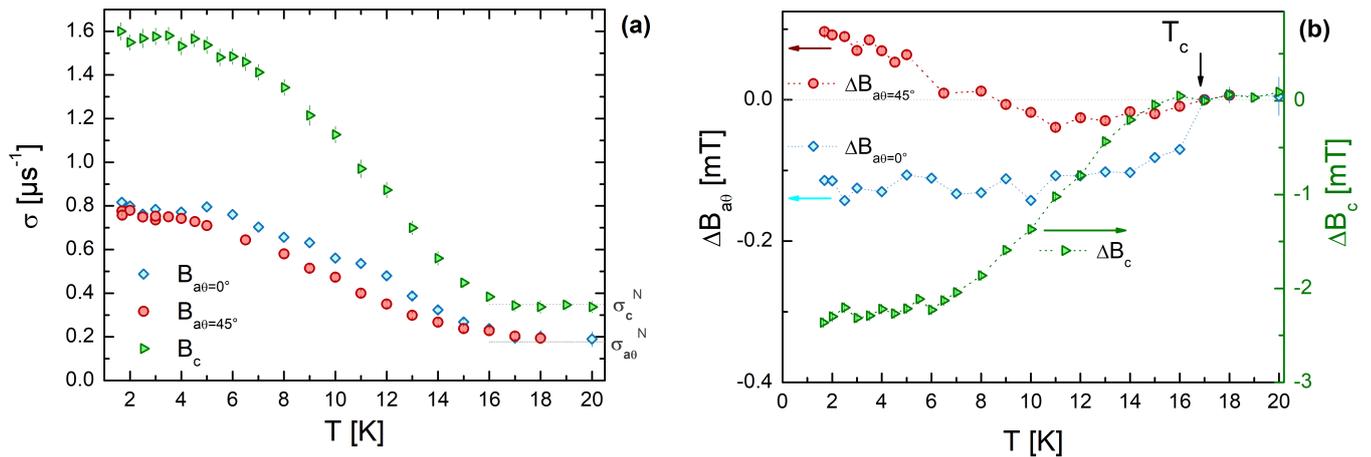

FIG. 3. **µSR relaxation — in-plane vs. out-of-plane results.** Temperature evolution of the fit parameters, as resulting from fits of the experimental asymmetry in the time domain [25]: (a) TF-µSR depolarization rate in a magnetic field applied in the *ab*-plane ($B_{a0}$, $B_{a45}$) and out-of-plane ($B_c$). (b) Diamagnetic field shift recorded in all the applied-field configurations.

This orbital richness provides a foundation for multiorbital pairing channels, opening up possibilities beyond the conventional $s^\pm$ spin singlet, and enabling both intra- and inter-orbital SC pairing in spin-triplet channels [33]. In this scenario, spin-orbit coupling (SOC) plays a significant role. Indeed, in LiFeAs, SOC is not merely an atomic $\bm{L}\cdot\bm{S}$ term, but rather a multiorbital interaction. Projecting it onto the dominant ($d_{xz}, d_{yz}$) subspace yields an effective form proportional to $L_z S_z$. Additional terms of comparable magnitude arise from coupling with other orbitals and involve spin components in the plane [33]. Since LiFeAs is centrosymmetric, time-reversal and inversion symmetry guarantee the Kramer degeneracy of its normal-state quasiparticle spectrum. Consequently, each band is doubly degenerate, but the Kramers pairs generally have different orbital characters due to SOC.

In a single-band description, SC pairing is energetically favored when it connects states within the same band. Thus, by rewriting the SC pairing in the band basis that diagonalizes the SOC, it is possible to establish the most favoured direction of the $\bm{d}$ vector in the presence of SOC.

In the case of *intra*-orbital pairing, only the $d_z$ component remains purely intraband in the band basis, while the $d_x$ and $d_y$ components acquire a significant interband character and are thus suppressed by SOC, effectively pinning the $\bm{d}$ vector along the $z$ axis [25].
In the presence of SOC, the normal–state eigenstates contain coherent superpositions of $d_{xz}$ and $d_{yz}$ states, thus pairing between the two orbitals is expected to occur without a large energy cost. Therefore, *Inter*–orbital pairing, which in the absence of SOC is allowed by symmetry, but it may be subleading, for relatively strong SOC can be the dominant pairing channel. Its symmetry with respect to the orbital degree of freedom then determines the energetically favored direction of the $\bm{d}$-vector. In the *antisymmetric inter-orbital channel*, the overall antisymmetry of the Cooper pairs requires an even-parity momentum structure. Like for the inter–orbital pairing, only the $d_z$ component is purely intraband, so that the $\bm{d}$-vector assumes the preferred orientation along the $z$-axis. However, if additional terms beyond the minimal $L_z S_z$ SOC are included, like a nonlocal SOC or orbital hybridization, the $d$ vector can develop non vanishing $d_x, d_y$ components [33]).

On the other side, the *symmetric inter-orbital channel*, odd in momentum, gives only purely intraband planar triplet components ($d_x, d_y$), being $d_z$ energetically suppressed by SOC due to its interband character. This pairing corresponds to an energetically stable multicomponent order parameter, with finite $d_x$ and $d_y$ components, compatible with the conditions required for the emergence of CLV.

Among the possible odd-parity form factors compatible with the tetragonal symmetry and the nematic SC character of LiFeAs, we identify two primary nematic candidates. The first one is $\bm{d}(\bm{k}) = \Delta_0(\sin k_z, \sin k_z, 0)$. Although $\sin k_z$ has zeros at $k_z = 0$ and $\pi$ within the Brillouin zone, these zeros do not necessarily intersect the actual Fermi surfaces of LiFeAs. Thus, this gap factor can produce a *nodeless superconducting state*. An alternative nodeless, odd-parity momentum structure can be found by considering the linear combination of the $E_u$ components $\bm{d}_1 = (\sin k_x \cos k_y, 0, 0)$ and $\bm{d}_2 = (0, \sin k_y \cos k_x, 0)$. In this second case, a fully symmetric combination $\bm{d}_1 + \bm{d}_2$ would preserve the tetragonal invariance. However, the spontaneous selection of a single component, or an unequal weighting of the two, generates a nematic phase. This state can be fully gapped on the LiFeAs Fermi surfaces [25].

Therefore, the strong $xz/yz$ hybridization and SOC in LiFeAs can stabilize a unitary, nodeless state with an in-plane $\bm{d}$-vector. This can result in a nematic triplet with a fully-gapped order parameter, consistent with both the experi-mentally observed saturated behavior of the depolarization rate ($\sigma \propto n_s$, the superfluid density) and with the minimal condition for the occurrence of nontrivial vortex textures, including skyrmion-like vortices [6].

*Evolution of peak splitting with temperature*

We model a nematic superconductor using the two-component Ginzburg-Landau model from Refs. [7, 8, 34]. The multidimensional OP exhibits an *in*-plane $\boldsymbol{d}$-vector (director) and *out-of*-plane applied field $\boldsymbol{B}_{\text{app}}$. The Gibbs free energy is given by,

$$G = \int_\Omega \left\{ \frac{1}{2} Q_{ij}^{\alpha\beta} \overline{D_i \psi_\alpha} D_j \psi_\beta + \frac{1}{2} \left| \boldsymbol{B}_{\text{loc}} - \boldsymbol{B}_{\text{app}} \right|^2 + F_p(|\psi_\alpha|, \theta_{\alpha\beta}) \right\}, \quad (2)$$

where $D_i = \partial_i - iA_i$ is the covariant derivative, $\boldsymbol{A}$ is the vector potential, $\boldsymbol{B}_{\text{loc}} = \nabla \times \boldsymbol{A}$ the corresponding local magnetic field and $\boldsymbol{B}_{\text{app}}$ the applied magnetic field. The two superconducting OPs are complex fields $\psi_\alpha = |\psi_\alpha| e^{i\varphi_\alpha}$, $\alpha \in \{1, 2\}$. The potential (non-gradient) terms $F_p$ and gradient couplings $Q_{ij}^{\alpha\beta}$ are discussed in detail in the SM [25].

The multiple coupled components introduce additional characteristic length scales and anisotropies arising from hybridised collective modes [35]. These qualitatively modify inter-vortex interactions and consequently the mixed state [5]. The nematic model goes beyond this, modifying the interactions of individual fractional vortices that make up a single full vortex. This is a result of the cross-terms, which favour non-coincident zeroes of the two condensates ($\psi_{1,2} = 0$). Hence, the complex vector order parameter $\Psi = (\psi_1, \psi_2)$ never vanishes and a skyrmion or coreless vortex is formed (see Fig. 1).

Skyrmions can experimentally appear only in a modified mixed state. Hence, we use an approach presented in Ref. [7] and generalized in detail in Ref. [36] to find lattice solutions without assuming symmetries. A periodic (lattice) solution, is a minimiser of the Gibbs free energy per unit volume, $\langle G \rangle = G(\Omega)[\Phi, A]/|\Omega|$ where $\Omega$ is a periodic unit cell and $|\Omega|$ it's volume. We use a gradient decent method to minimise $\langle G \rangle$ w.r.t. the fields $(\Psi, A)$ while fixing $\Omega$, and algebraically minimising $\langle G \rangle$ w.r.t. the lattice cell geometry $\Omega$ while fixing the fields $(\Psi, A)$.

We consider an applied field of $H = 1.5 H_{c_1}$, parallel to the $c$-axis. Then, for each value of the temperature $T$, we find the minimal energy lattice, shown in Fig. 4. The final panel is the magnetic field probability density, calculated as a normalized sum of localized Gaussians:

$$p(B) = \frac{1}{hn^2} \sum_{i=1}^{n} \sum_{j=1}^{n} \frac{1}{\sqrt{2\pi}} e^{-\frac{1}{2} \left( \frac{B - B_{ij}}{h} \right)^2}, \quad (3)$$

where $B_{ij}$ is the local magnetic field at the lattice site $(i, j)$, on a discretised grid of $n^2$ lattice points. The parameter $h$ determines the smoothing and should be set relative to the number of grid points $n^2$.

There are three possible phases for the results:

- *Low-T phase* – fractional vortices split to form skyrmions that are well separated. This dilute phase will be dominated by other effects (e.g. pinning) as the inter-vortex forces are so weak. While strictly different from the Abrikosov lattice, the probability density will not have a resolvable double peak due to the large difference between fractional vortex separation and skyrmion separation.

- *Mid-T phase* – Skyrmions have directional interactions, thus forming chains, the direction of which is determined by the fractional vortex-splitting direction. The probability density now features a clear double peak, where each peak corresponds to a different length scale, the inter-chain separation and the chain-link length. As temperature increases the inter-chain separation, and hence peak separation, decreases.

- *High-T phase* – The vortex pressure is sufficiently large to force the chains to merge, causing the fractional vortices to combine into composite ones tightly packed into an AVL. The probability density shows again a single peak.

Not all parameter choices in our GL model will exhibit all of the above three phases. If all three are present as temperature increases, we observe a single peak become a double peak, then the two peaks merge into a single peak. The parameter choice we make exhibit the second and third phases, demonstrating the clear double peak phase. The peaks come together and eventually merge into an AVL as shown in Fig. 2B-(f).

## IV. CONCLUSION

Thanks to *angle-resolved* TF-µSR, for the first time, we were able to provide clear evidence of the presence and temperature evolution of CLV in a bulk nematic superconductor. This result is significant as it suggests that the standard Brandt model [37, 38], normally used to extract the magnetic penetration depth from the measured superconducting depolarization rates, is unsuitable for unconventional vortex lattices, thus requiring a new scaling law, relating $\sigma$ and $\lambda$.

**Data availability**

All the data needed to evaluate the reported conclusions are presented in the paper and/or in the supplementary material [25]. Additional data related to this paper may be requested from the authors. The µSR data were generated at the SµS (Paul Scherrer Institut, Switzerland). Derived data supporting the results of this study are available from the corresponding authors or beamline scientists. The `musrfit` software package is available online free of charge at https://lmu.pages.psi.ch/musrfit/.

**Acknowledgements**

We acknowledge the allocation of beam time at the Swiss muon source (VMS µSR spectrometer). We kindly acknowledge I. Maccari and E. Babaev for fruitful discussions. PG acknowledges financial support from the Italian Ministry of University and Research (MUR) under the National Recovery and Resilience Plan (NRRP), Call PRIN 2022, funded by the European Union NextGenerationEU, PNRR-M4C2, I1.1 project 2022HTPC2B (TOTEM)- CUP B53D23004210006 and PNRR-M4C2, I1.3 Extended Partnership NQSTI - PE0000023 CUP B53C22004180005. TW acknowledges the University of Edinburgh for funding a postdoctoral position for the majority of this work. SW and BB acknowledges financial support from the

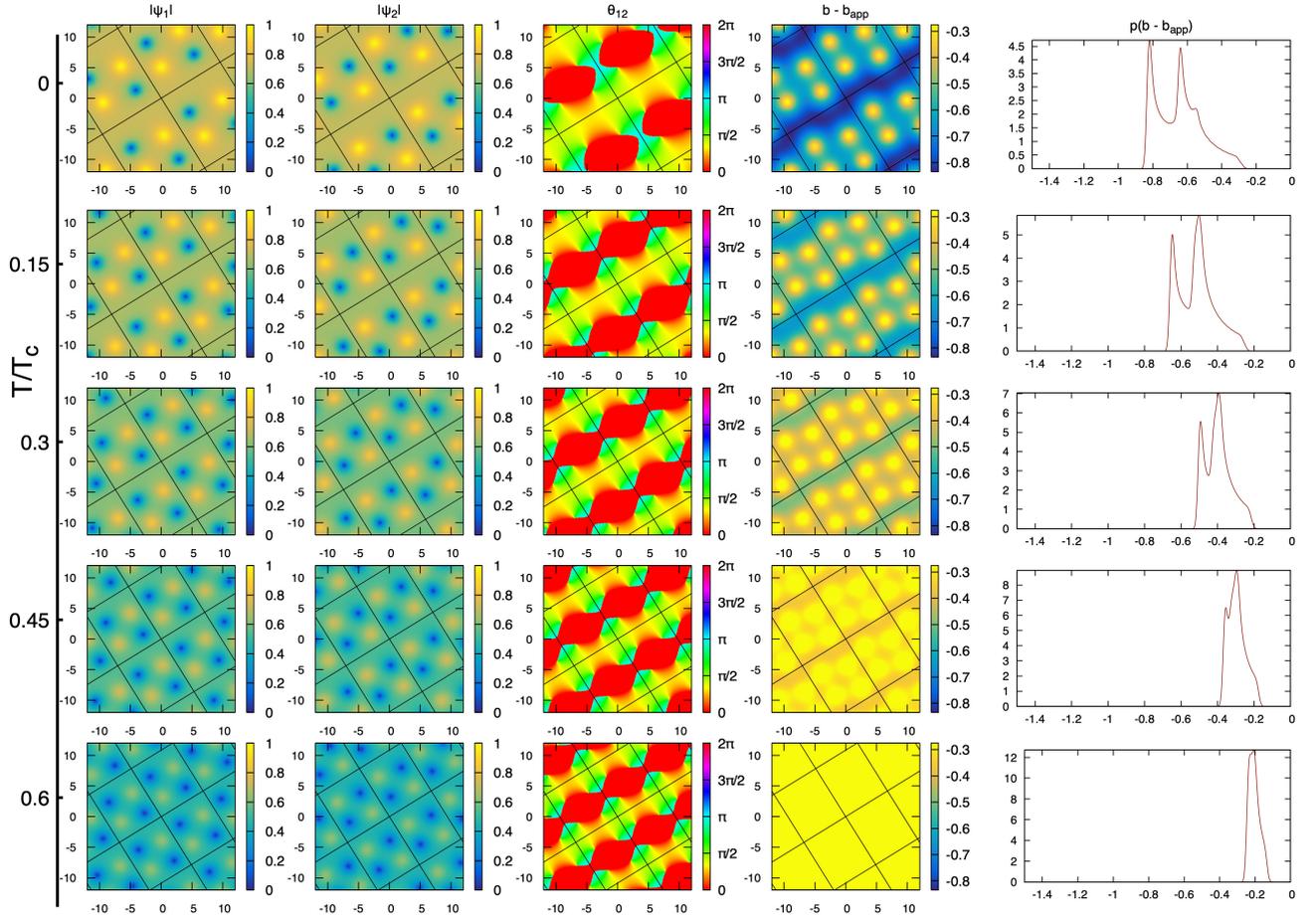

FIG. 4. **Contour plots of the SC skyrmion chains** (i.e., lattice solutions) for the model given by Eq. (2). From left to right: OP magnitudes $|\psi_1|$ and $|\psi_2|$, relative phase $\theta_{12}$, local magnetic field strength $\boldsymbol{B}_{\text{loc}}$, and its probability density $p(B)$.

**Author contributions**

μSR experiments and data analysis: G.L., T.S. DC-magnetometry experiments and data analysis: G.L. Skyrmionic vortex lattice theory: T.W., M.S. Multidimensional order parameter theory: P.G. Sample growth and structural characterization: F. A, S. W., B. B. The manuscript was written by G.L., T.S., P.G., T.W. with input from all authors. Project planning and coordination: G.L.

**Competing interests**

The authors declare no competing financial or non-financial interests.

**Correspondence** and requests for materials should be addressed to G.L., T.S., P.G., T.W.

## References

\* Corresponding author: gianrico.lamura@spin.cnr.it
† Corresponding author: twinyard001@dundee.ac.uk
‡ Corresponding author: paola.gentile@spin.cnr.it
§ Corresponding author: tshiroka@phys.ethz.ch

# Supplementary Material for the manuscript
# "Unconventional mixed state in the nematic superconductor LiFeAs"

G. Lamura,[1,*] T. Winyard,[2,3,†] P. Gentile,[4,‡] M. Speight,[5] F. Anger,[6] B. Büchner,[6] S. Wurmehl,[6] and T. Shiroka[7,8,§]

[1]*CNR-SPIN, I-16152 Genova, Italy*
[2]*Division of Mathematics, University of Dundee, Dundee DD1 4HN, United Kingdom*
[3]*Maxwell Institute, University of Edinburgh, Edinburgh EH9 3FD, United Kingdom*
[4]*CNR-SPIN, c/o Universitá degli Studi di Salerno, I-84084 Fisciano (SA), Italy*
[5]*School of Mathematics, University of Leeds, Leeds LS2 9JT , United Kingdom*
[6]*Leibniz-Institute for Solid State and Materials Research, IFW-Dresden, 01069 Dresden, Germany*
[7]*Laboratory for Muon-Spin Spectroscopy, Paul Scherrer Institut, CH-5232 Villigen PSI, Switzerland*
[8]*Laboratorium für Festkörperphysik, ETH Zürich, CH-8093 Zürich, Switzerland*

## Materials and Methods

### I. SAMPLE SYNTHESIS

Platelet-like single crystals of stoichiometric LiFeAs were grown using the self-flux method as described in Ref. [1]. LiFeAs crystallizes in the tetragonal $Cu_2Sb$/PbClF structure type, belonging to the $P4/nmm$ (No. 129) space group, with cell parameters $a = b = 3.7678(4)$ Å and $c = 6.3151(8)$ Å. A schematic view of the crystal structure is shown in Fig. 1d of the main article.

### II. DC-MAGNETIC SUSCEPTIBILITY

Complementary dc-susceptibility measurements were performed after the µSR experiments employing a commercial MPMS2 SQUID magnetometer by Quantum Design. We used a parallel-field configuration to minimize the demagnetizing effects. Figure SI-1 shows the low-field dc-magnetic susceptibility $\chi$ measured in an external magnetic field of 1 mT using the ZFC protocol. The full shielding ($\chi \sim 0.83$) at the lowest temperature is indicative of bulk superconductivity.

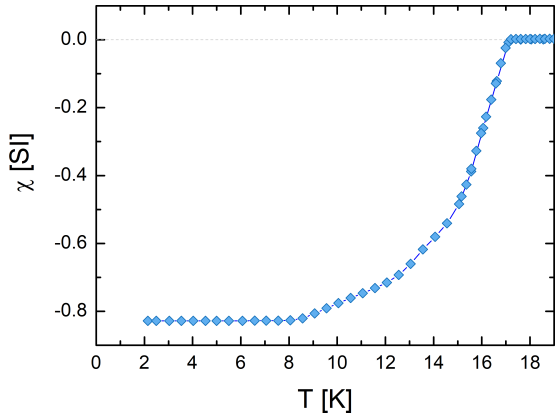

Fig. SI-1. **Dc-Susceptibility**. Temperature-dependent magnetic susceptibility of LiFeAs measured in an applied field of 1 mT. The ZFC protocol, with the magnetic field applied parallel to the $ab$-plane, was used.

### III. MUON-SPIN ROTATION AND RELAXATION ( µSR)

The bulk µSR measurements were carried out at the versatile muon spectrometer (VMS) on the πE1 beamline of the Swiss muon source at Paul Scherrer Institut, Villigen, Switzerland. The sample was mounted on a sample holder consisting of a copper frame holding a 25-µm thick copper foil. Apiezon-N grease was used to ensure good thermalization at low temperatures. Since LiFeAs is highly sensitive to air, the mounting was performed in an Ar-atmosphere glove box. The sample was then quickly inserted into the evacuated VMS cryostat and cooled down using a few millibars of He exchange gas.

#### A. In-plane TF-µSR measurements

The time-dependent muon-spin polarization for the two in-plane field configurations shown in Fig. 2a was fitted by the superposition of three Gaussian distributions:

$$P_{\text{TF}}(t) = \sum_{i=1}^{2} \frac{A_i}{A_0} \cos(\gamma_\mu B_i t + \phi) e^{-\sigma_i^2 t^2/2} + \\ + \frac{A_{\text{bg}}}{A_0} \cos(\gamma_\mu B_{\text{bg}} t + \phi) e^{-\sigma_{\text{bg}}^2 t^2/2}. \quad (1)$$

Here, $A_0$ is the total asymmetry, while $A_{i=1,2}$ are the asymmetries of the two Gaussian components describing the superconducting volume fraction. $\sigma_{i=1,2}$ and $B_{i=1,2}$ represent the corresponding Gaussian relaxation rates and the local field sensed by the muons in the vortex state. $A_{\text{bg}}$ is the contribution of those muons implanted in the non superconducting parts of the sample (or parts of the sample holder) precessing around the field $B_{\text{bg}} = B_{\text{app}}$, with $\sigma_{\text{bg}}$ being the depolarization rate. Here, $A_{\text{bg}}$, $B_{\text{bg}}$, and $\sigma_{\text{bg}}$ were estimated in the $\boldsymbol{B}_\parallel$ configuration (see next section) where, due to the faster decay, only the background contribution remains at long times. Such values were then kept fixed in all the applied field configurations. $\gamma_\mu/2\pi = 135.53$ MHz/T is the muon gyromagnetic ratio and $\phi$ is a shared initial phase. $\sigma_{i=1,2}$ is temperature-independent in the normal state but, below $T_c$, it starts to increase due to the onset of the FLL. Without any additional hypotheses about the nature of the vortex lattice, it is possible to extract the resulting depolarization rate $\sigma$ from the second moment of the local field

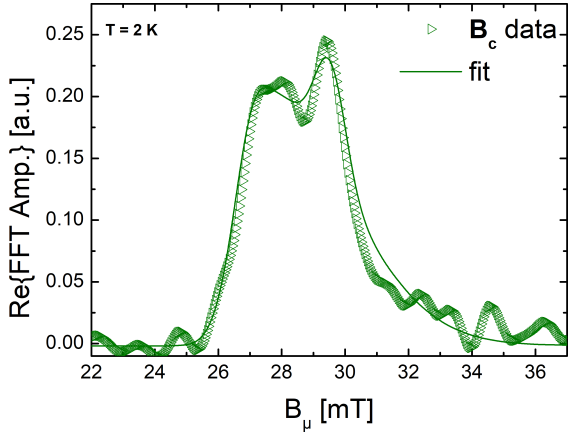

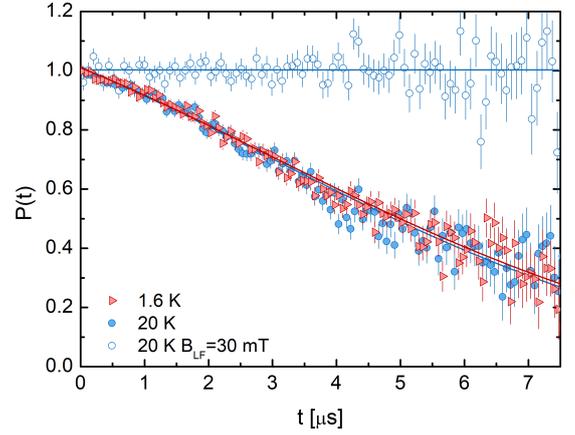

Fig. SI-2. **TF-μSR FFT SkG model fit.** FFT of the TF-μSR Asymmtry with a weak apodization recorded at 2K in a magnetic field applied parallel to the $c$-axis ($B_c$). The continuous line represents the SkG model fit.

Fig. SI-3. **ZF-μSR.** Time-dependent ZF-μSR polarization collected in the superconducting (1.6 K) and in the normal state (20 K) of LiFeAs. The essentially overlapping datasets indicate the absence of TRS breaking, whose occurrence would have resulted in a stronger decay below $T_c$. Void symbols represent a muon-spin decoupling measurement in an applied LF-field of 30 mT.

distribution $p(B)$ from the equation [2]:

$$\sigma^2 = \sum_{i=1}^{2} A_i [\sigma_i^2 - (B_i - \langle B \rangle)^2 \gamma_\mu^2]/(A_1 + A_2), \quad (2)$$

where $\langle B \rangle = (A_1 B_1 + A_2 B_2)/(A_1 + A_2)$.

Considering that the nuclear dipolar contribution to the muon-spin relaxation rate $\sigma_n$ is essentially $T$-independent in the 0–20 K range, the relaxation rate in the SC state, $\sigma_{sc}$, can be extracted from the total depolarization rate $\sigma$ via a quadrature subtraction:

$$\sigma_{sc} = \sqrt{\sigma^2 - \sigma_n^2}. \quad (3)$$

### B. Out-of-plane TF-μSR measurements

The time-dependent muon-spin polarization for the out-of-plane applied-field orientation (Fig. 2c) shows a fast depolarization. Its FFT (Fig. 2e) exhibits a clear peak splitting, visible up to 6 K (Fig. 2f). Such feature makes the superconducting peak particularly asymmetric. The best fit of the time dependent polarization was then obtained by the sum of a damped Gaussian component (for the background contribution) and a skewed Gaussian function, skG($t$) [3], for the superconducting component, since its FT consists of an asymmetric field distribution [4]:

$$P_{TF}(t) = \frac{A_{skG}}{A_0} \cdot \text{skG}(B_0, \sigma_+, \sigma_-, t) + \frac{A_{bg}}{A_0} \cos(\gamma_\mu B_{bg} t + \phi) \cdot e^{-\sigma_{bg}^2 t^2/2}. \quad (4)$$

Here, $B_0$, $\sigma_+$, and $\sigma_-$ are the maximum of $p(B)$ distribution and its lower and higher limits. Figure SI-2 shows the FFT of the polarization, recorded at 2 K, and of the SkG model, which fits fairly well the experimental data.
Since $P(t)$ decays faster in the **B** ∥ **c** configuration, the background contribution remains clearly detectable at long times. Thus the parameters of the background contribution ($A_{bg}$, $B_{bg}$, and $\sigma_{bg}$) could be determined by fitting the long time tail of $P(t)$ ($t > 5\,\mu$s). Particularly relevant is the evaluation of $A_{bg}$, as it provides the lower limit of the superconducting volume fraction: $V_{sc}(2\,\text{K}) = 1 - A_{bg}/A_{tot} \simeq 89\,(1)\,\%$. This is consistent with the superconducting shielding fraction estimated by DC susceptibility (see section II).
The first and second moments of the skewed Gaussian distribution are:

$$\langle B \rangle = B_0 + \sqrt{\frac{2}{\pi}} \frac{\sigma_+ - \sigma_-}{\gamma_\mu} \quad (5)$$

$$\langle \Delta B^2 \rangle = \frac{1}{\pi \gamma_\mu^2} \cdot \left[ (\pi - 2)\sigma_-^2 - (\pi - 4)\sigma_+ \sigma_- + (\pi - 2)\sigma_+^2 \right].$$

The temperature evolution of the resulting depolarization rate $\sigma_c$ and of the diamagnetic field shift $\Delta B_c$ are reported in Fig. 3).

### C. ZF-μSR measurements

We performed zero-field ZF-μSR measurements in the normal- and superconducting states in order to assess the possible occurrence of time reversal symmetry (TRS) breaking in the superconducting phase. This technique is highly sensitive to weak spontaneous fields, which manifest as an additional relaxation rate below $T_c$. ZF-μSR is among the few techniques (the other one being polar Kerr effect) which allows to detect TRS breaking, as observed, e.g., in Sr$_2$RuO$_4$ [5–7], in Re-based superconductors [8–12], or in the putative topological superconductor Sr$_{0.1}$Bi$_2$Se$_3$ [13]. Figure SI-3 shows the time-dependent ZF-μSR muon spin polarization collected above- (20 K) and below $T_c$ (1.6 K), which overlap almost fully. Further, the ZF relaxation is promptly recovered by the application of a small external magnetic field parallel to the muon spin polarization, $B_{LF} = 30$ mT (Fig. SI-3) (here, LF denotes longitudinal field). The combined evidence from LF- and ZF-μSR experiments demonstrates that the measured ZF-μSR relaxation is only due to static local fields at the microsecond timescale. Both spectra were fitted using the Gaussian Kubo-Toyabe (GKT) depolarization function, which is suited for modelling local field distributions due to nuclear dipolar moments. An additional exponential term, $e^{-\Lambda_{ZF}t}$, was included to better account for the effects arising from dipolar fields of electronic origin [14]:

$$P_{ZF}(t) = \left[\frac{1}{3} + \frac{2}{3}(1 - \sigma_{ZF}^2 t^2)e^{-\frac{\sigma_{ZF}^2 t^2}{2}}\right] \cdot e^{-\Lambda_{ZF}t}. \quad (6)$$

Here, $\sigma_{ZF}$ and $\Lambda_{ZF}$ represent the zero-field Gaussian and Lorentzian relaxation rates, respectively. The solid lines in Fig. SI-3 are fits to the above equation, yielding $\sigma_{ZF} = 0.106(6)\,\mu s^{-1}$ and $\Lambda_{ZF} = 0.086(5)\,\mu s^{-1}$ at 20 K and $\sigma_{ZF} = 0.104(5)\,\mu s^{-1}$ and $\Lambda_{ZF} = 0.085(5)\,\mu s^{-1}$ at 1.6 K, respectively. The relaxation rates in the normal- and the superconducting states are almost identical confirming the absence of any additional μSR relaxation below $T_c$, thus definitively ruling out the breaking of TRS in the superconducting state of LiFeAs and, therefore, also the presence of either chiral or nonunitary spin-triplet states.

### D. Superfluid density and superconducting gap

To estimate the magnetic penetration depth and the superconducting gap we approximate the skyrmionic mixed SC state with a standard Abrikosov vortex lattice. Thus, for applied fields sufficiently low with respect to the upper critical field, the effective magnetic penetration depth can be extracted from the measured $\sigma_{sc}$ by using the Brandt's scaling equation [15, 16]:

$$\frac{\sigma_{sc}^2(T)}{\gamma_\mu^2} = 0.00371 \frac{\Phi_0^2}{\lambda_{eff}^4(T)}, \quad (7)$$

where $\Phi_0 = 2.07 \times 10^3\,\text{T nm}^2$ is the magnetic flux quantum. In our case, the applied magnetic field intensity $B_{app} = 14.5\,\text{mT}$ (or 30 mT) is significantly lower than the reported upper critical field $\mu_0 H_{c2}^c(0) \sim 15(1)\,\text{T}$ and $\mu_0 H_{ab}^\perp(0) \sim 24(4)\,\text{T}$ [17, 18]. Hence, we can ignore the effects of the overlapping vortex cores. We note also that such applied fields are at the verge of the lower critical fields, as reported in the literature [18]. Since the $H_{c1}$ value may depend on the sample quality, measurement conditions, and the definition of $H_{c1}$ itself, it is not unusual that, as in our case, the in-plane lower critical field at 5 K to be slightly higher than the values reported in the literature. We verified that in our TF-μSR measurements, with the applied magnetic field parallel to the sample surface, there was a small volume fraction fully in the Meissner state. We estimated its fraction to be less than 1.5%, i.e., it is absolutely negligible.

Figures SI-4a-b show the temperature-dependent inverse-square of the magnetic penetration depth ($\propto n_S$). Since the curves saturate at low temperature and no inflection points are visible, in all the three field-direction cases we could satisfactorily fit the experimental data by a single $s$-wave gap model:

$$\frac{\lambda^2(0)}{\lambda^2(T)} = \left[1 - 2\int_{\Delta(0)}^\infty \left(-\frac{\partial f}{\partial E}\right) \cdot \frac{E\,dE}{\sqrt{E^2 - \Delta^2(T)}}\right], \quad (8)$$

where $f = (1 + e^{E/k_B T})^{-1}$ is the Fermi function and $\Delta(T) = \Delta_0 \tanh\left[1.82[1.018(T_c/T - 1)]^{0.51}\right]$ [19, 20], with $\Delta_0$ being the SC gap value at 0 K. Here, $\lambda(0)$ and $\Delta_0$ are the zero-temperature limit of the penetration depth and the superconducting gap, respectively. All the fit parameters are summarized in Table I.

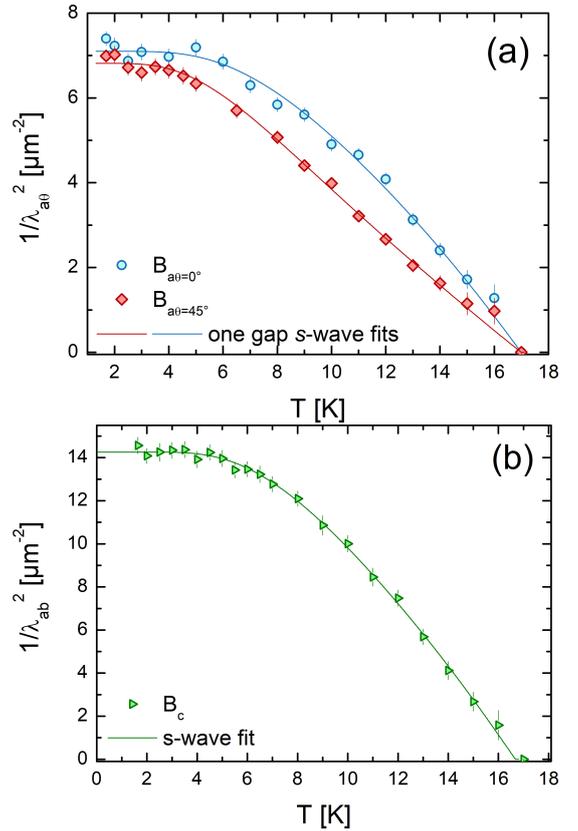

Fig. SI-4. **Magnetic penetration depth.** Temperature dependence of the inverse square of the magnetic penetration depth as resulting from Eq. (7): (a) $\lambda_{ac}$ ($B_{a0}$ and $B_{a45}$), b) $\lambda_{ab}$ ($B_c$). Continuous lines represent fits using the model given by Eq. (8).

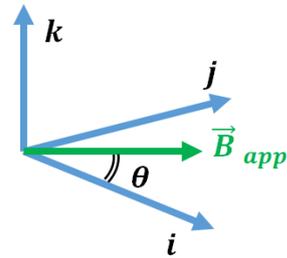

Fig. SI-5. **Applied magnetic field direction.** Diagram illustrating the direction of the applied magnetic field $B_{app}$ with respect to a generic $ijk$ reference system.

Table I. The $\lambda(0)$ and $\Delta_0$ parameters (referring to 0-K values), as resulting from a fit to Eq. (8).

| $B_{app}$ | $\lambda(0)$ [nm] | $T_c$ [K] | $\Delta_0$ [meV] |
|---|---|---|---|
| $B_{a0} = 14.5\,\text{mT}$ | 375 (2) | 17.0 (3) | 2.57 (5) |
| $B_{a45} = 14.5\,\text{mT}$ | 383 (2) | 17.0 (3) | 2.01 (4) |
| $B_c = 30\,\text{mT}$ | 265 (1) | 16.7 (3) | 2.46 (7) |

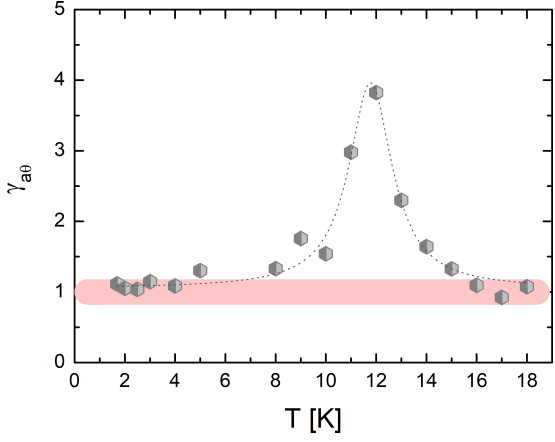

Fig. SI-6. **In-plane anisotropy**. Temperature dependence of the anisotropy factor $\gamma_{a\theta}$ for the *in*-plane applied magnetic field $B_{a45}$.

### E. Anisotropic superconductivity in LiFeAs

Decades ago, the angular variation of the second moment of the internal field distribution, $p(B)$, was calculated with respect to the direction of the applied magnetic field under the hypothesis of a conventional vortex lattice. This was done in order to disentangle the contributions of each field penetration depth in orthorhombic YbBa$_2$Cu$_3$O$_{7-\delta}$ [21–23]. For a magnetic field oriented along a generic direction relative to the crystalline axes, the second moment can be modelled using the equation:

$$\langle \Delta B^2 \rangle^{1/2}(T,\theta) = \langle \Delta B^2 \rangle^{1/2}(T,0)\left[\cos^2\theta + \frac{1}{\gamma_{ij}^2}\sin^2\theta\right]^{1/2} \quad (9)$$

where $\gamma_{ij} = \sqrt{M_j/M_i} = \lambda_i/\lambda_j$. $M_i$ is the effective mass of an electron flowing along the *i*th principal axis and $\lambda_i$ is the penetration depth of magnetic fields applied along the same axis. $\theta$ represents the angle between the applied field and the *i*th axis for rotations about the *k*th axis (see Fig. SI-5).

When applied to the $B_{a0}$ and $B_{a45}$ datasets, Eq. (9) allows us to extract the in-plane anisotropy factor $\gamma_{a\theta}$, whose temperature behavior is reported in Fig. SI-6. Note that, $\gamma_{a\theta} \sim 1$ at low temperatures and near $T_c$, while it presents a broad peak in the 8–16 K range. This temperature range is just above the paramagnetic upturn measured in the diamagnetic shift for an applied field $B_{a45}$ (see main article). Although obtained under *the oversimplified assumption* of a conventional vortex lattice, the behavior of $\gamma_{a\theta}$ clearly suggests a breaking of rotational symmetry for in-plane applied magnetic fields. Indeed, for symmetry reasons, in a tetragonal system the second moment of the local field distribution is expected to be invariant, i.e., independent of the direction of the applied magnetic field in the *ab*-plane, as stated by Eq. (9).

## IV. MINIMAL TWO–ORBITAL MODEL FOR LIFEAS AND THE ROLE OF ATOMIC SPIN–ORBIT COUPLING

We consider a minimal two–orbital model for the Fe $d_{xz}$ and $d_{yz}$ states, following the construction reported in Ref. [24]. We work in the orbital ⊗ spin basis

$$\Psi_{\mathbf{k}} = \begin{pmatrix} c_{xz,\uparrow}(\mathbf{k}) \\ c_{xz,\downarrow}(\mathbf{k}) \\ c_{yz,\uparrow}(\mathbf{k}) \\ c_{yz,\downarrow}(\mathbf{k}) \end{pmatrix}. \quad (10)$$

In this basis, the normal–state Hamiltonian can be written as

$$H_0(\mathbf{k}) = H_{\text{kin}}(\mathbf{k}) + H_{\text{SOC}}, \quad (11)$$

with

$$H_{\text{kin}}(\mathbf{k}) = \epsilon_0(\mathbf{k})\,\tau_0 \otimes \sigma_0 + \gamma(\mathbf{k})\,\tau_x \otimes \sigma_0, \quad (12)$$

where $\tau_i$ and $\sigma_i$ are the Pauli matrices in orbital- and spin space, respectively, $\epsilon_0(\mathbf{k})$ is the average dispersion, and $\gamma(\mathbf{k})$ encodes the orbital hybridization due to the crystal field and hopping anisotropy. The key ingredient of the model is the atomic spin–orbit coupling (SOC),

$$H_{\text{SOC}} = \lambda\,\mathbf{L}\cdot\mathbf{S} \approx \lambda\,\tau_y \otimes \sigma_z, \quad (13)$$

which is the leading SOC term in the $(d_{xz}, d_{yz})$ subspace. This operator mixes the two orbitals in such a way that

$$H_{\text{SOC}}|d_{xz}\uparrow\rangle \propto |d_{yz}\uparrow\rangle, \qquad H_{\text{SOC}}|d_{xz}\downarrow\rangle \propto -|d_{yz}\downarrow\rangle, \quad (14)$$

and similarly for the other orbital. As a consequence, the eigenstates of $H_0(\mathbf{k})$ are Kramers pairs with mixed orbital character. The presence of SOC has a direct impact on the energy levels of different triplet pairing channels. A generic triplet gap matrix in the orbital basis can be written as

$$\hat{\Delta}(\mathbf{k}) = (\mathbf{d}(\mathbf{k})\cdot\boldsymbol{\sigma})\,i\sigma_y \otimes \hat{\Phi}, \quad (15)$$

where $\hat{\Phi}$ is a matrix in orbital space. Two relevant possibilities are:

1. **Intra–orbital triplet**, which pairs electrons within the same orbital. In this case, the orbital structure is diagonal in orbital space and thus $\hat{\Phi} = \tau_0$, the identity in orbital space. Since the pairing is spin-triplet (thus even in the spin degree of freedom) and orbital-symmetric, Fermi statistics require the momentum dependence of the gap function to be *odd* to guarantee its overall antisymmetry under the exchange of the two electrons forming the Cooper pair. As a consequence $d(\mathbf{k})$ is an odd function of $\mathbf{k}$:

$$\mathbf{d}(-\mathbf{k}) = -\mathbf{d}(\mathbf{k}). \quad (16)$$

2. Inter–orbital triplet which pairs electrons in different orbitals. There are two different possibilities:
(i) **Interorbital triplet pairing with odd symmetry in orbital space (OO)**, such that $\hat{\Phi} = \tau_y$:

$$\Delta^{OO}(\mathbf{k}) = (\mathbf{d}^{\text{even}}(\mathbf{k})\cdot\boldsymbol{\sigma})\,i\sigma_y\,\tau_y, \quad (17)$$

with $\mathbf{d}^{\text{even}}(-\mathbf{k}) = \mathbf{d}^{\text{even}}(\mathbf{k})$, an even function of $\mathbf{k}$;
(ii) **Interorbital triplet pairing with even symmetry in orbital space (OE)**, which must be odd in k, such that $\hat{\Phi} = \tau_x$:

$$\Delta^{OE}(\mathbf{k}) = (\mathbf{d}^{\text{odd}}(\mathbf{k})\cdot\boldsymbol{\sigma})\,i\sigma_y\,\tau_x, \quad (18)$$

where $\mathbf{d}^{\text{odd}}(-\mathbf{k}) = -\mathbf{d}^{\text{odd}}(\mathbf{k})$.

We can now provide a direct energetic argument for the stability of the superconducting **d**-vector in the presence of spin–orbit coupling. The argument is formulated in the basis that diagonalizes the spin–orbit interaction (band-basis) and relies on the degeneracy structure of the normal-state eigenstates. We consider the expression assumed by each components of the intra-orbital and inter-orbital triplet pairings in the band basis that diagonalizes the SOC term. Since it depends on $\tau_y \otimes \sigma_z$ and being $\sigma_z$ diagonal in the considered orbital $\otimes$ spin basis, we can diagonalize the orbital part of the SOC term separately for each spin projection. The matrix $\tau_y$ has eigenvalues $\pm 1$ with eigenvectors (corresponding to the eigenvalues of the $z$ component of the angular orbital momentum $l_z$) given by:

$$|+\rangle = -\frac{1}{\sqrt{2}}(|xz\rangle + i|yz\rangle), \qquad |-\rangle = \frac{1}{\sqrt{2}}(|xz\rangle - i|yz\rangle). \tag{19}$$

Thus, the SOC eigenstates are

$$\begin{aligned} |+,\uparrow\rangle: & \quad E = +\lambda, \\ |-,\uparrow\rangle: & \quad E = -\lambda, \\ |+,\downarrow\rangle: & \quad E = -\lambda, \\ |-,\downarrow\rangle: & \quad E = +\lambda. \end{aligned} \tag{20}$$

We now define the operators $a$ for the SOC eigenstates with eigenvalue $+E$, and $b$ for the eigenstates corresponding to the SOC eigenvalue $-E$ as:

$$\begin{aligned} a_\uparrow &= -\frac{1}{\sqrt{2}}\left(c_{xz,\uparrow} + i\, c_{yz,\uparrow}\right), \\ b_\uparrow &= \frac{1}{\sqrt{2}}\left(c_{xz,\uparrow} - i\, c_{yz,\uparrow}\right), \\ a_\downarrow &= \frac{1}{\sqrt{2}}\left(c_{xz,\downarrow} - i\, c_{yz,\downarrow}\right), \\ b_\downarrow &= -\frac{1}{\sqrt{2}}\left(c_{xz,\downarrow} + i\, c_{yz,\downarrow}\right). \end{aligned} \tag{21}$$

The inverse relations are

$$\begin{aligned} c_{xz,\uparrow} &= \frac{1}{\sqrt{2}}\left(b_\uparrow - a_\uparrow\right), \\ c_{yz,\uparrow} &= \frac{i}{\sqrt{2}}\left(a_\uparrow + b_\uparrow\right), \\ c_{xz,\downarrow} &= \frac{1}{\sqrt{2}}\left(a_\downarrow - b_\downarrow\right), \\ c_{yz,\downarrow} &= \frac{i}{\sqrt{2}}\left(a_\downarrow + b_\downarrow\right). \end{aligned} \tag{22}$$

### A. Intra-orbital pairing

Let us start by considering the intra-orbital $d_z$ component of the **d** vector, corresponding to opposite-spin pairing within the same orbital:

$$\Delta_{d_z}^{\text{intra}}(\mathbf{k}) \sim \sum_{\ell=xz,yz} \left( c_{\mathbf{k},\ell,\uparrow} c_{-\mathbf{k},\ell,\downarrow} + c_{\mathbf{k},\ell,\downarrow} c_{-\mathbf{k},\ell,\uparrow} \right). \tag{23}$$

Let us consider each orbital separately.

$$\begin{aligned} c_{\mathbf{k},xz,\uparrow} c_{-\mathbf{k},xz,\downarrow} &= \frac{1}{2}\left(b_{\mathbf{k}\uparrow} - a_{\mathbf{k}\uparrow}\right)\left(a_{-\mathbf{k}\downarrow} - b_{-\mathbf{k}\downarrow}\right) \\ &= \frac{1}{2}\left(b_{\mathbf{k}\uparrow} a_{-\mathbf{k}\downarrow} - a_{\mathbf{k}\uparrow} a_{-\mathbf{k}\downarrow} - b_{\mathbf{k}\uparrow} b_{-\mathbf{k}\downarrow} + a_{\mathbf{k}\uparrow} b_{-\mathbf{k}\downarrow}\right). \end{aligned} \tag{24}$$

Similarly,

$$\begin{aligned} c_{\mathbf{k},yz,\uparrow} c_{-\mathbf{k}yz,\downarrow} &= \left(\frac{i}{\sqrt{2}}\right)\left(\frac{i}{\sqrt{2}}\right)\left(a_{\mathbf{k}\uparrow} + b_{\mathbf{k}\uparrow}\right)\left(a_{-\mathbf{k}\downarrow} + b_{-\mathbf{k}\downarrow}\right) \\ &= -\frac{1}{2}\left(a_{\mathbf{k}\uparrow} a_{-\mathbf{k}\downarrow} + a_{\mathbf{k}\uparrow} b_{-\mathbf{k}\downarrow} + b_{\mathbf{k}\uparrow} a_{-\mathbf{k}\downarrow} + b_{\mathbf{k}\uparrow} b_{-\mathbf{k}\downarrow}\right). \end{aligned} \tag{25}$$

Summing both orbital contributions,

$$\left(c_{\mathbf{k},xz,\uparrow} c_{-\mathbf{k},xz,\downarrow} + c_{\mathbf{k},yz,\uparrow} c_{-\mathbf{k},yz,\downarrow}\right) \sim a_{\mathbf{k}\uparrow} a_{-\mathbf{k}\downarrow} + b_{\mathbf{k}\uparrow} b_{-\mathbf{k}\downarrow}. \tag{26}$$

and thus

$$\Delta_{d_z}^{\text{intra}}(\mathbf{k}) \sim \Delta_{d_z}^{\text{a}}(\mathbf{k}) + \Delta_{d_z}^{\text{b}}(\mathbf{k})$$

Thus, the intra-orbital $d_z$ component produces *pure intraband* pairing in the SOC basis.

We now consider the intra-orbital equal-spin triplet pairing, giving the $d_x$ and $d_y$ components of the **d** vector.
The $d_x$ component reads

$$\Delta_{d_x}^{\text{intra}}(\mathbf{k}) \sim -\sum_{\ell=xz,yz} \left( c_{\mathbf{k},\ell,\uparrow} c_{-\mathbf{k},\ell,\uparrow} - c_{\mathbf{k},\ell,\downarrow} c_{-\mathbf{k},\ell,\downarrow} \right). \tag{27}$$

The contribution coming from the pairing within the $d_{xz}$ orbital is

$$\begin{aligned} c_{\mathbf{k},xz,\uparrow} c_{-\mathbf{k},xz,\uparrow} &= \frac{1}{2}\left(b_{\mathbf{k},\uparrow} - a_{\mathbf{k},\uparrow}\right)\left(b_{-\mathbf{k},\uparrow} - a_{-\mathbf{k},\uparrow}\right) \\ &= \frac{1}{2}\left(b_{\mathbf{k},\uparrow} b_{-\mathbf{k},\uparrow} + a_{\mathbf{k},\uparrow} a_{-\mathbf{k},\uparrow} - b_{\mathbf{k},\uparrow} a_{-\mathbf{k},\uparrow} - a_{\mathbf{k},\uparrow} b_{-\mathbf{k},\uparrow}\right). \end{aligned} \tag{28}$$

Then, the contribution coming from the pairing within the $d_{yz}$ orbital is

$$\begin{aligned} c_{\mathbf{k},yz,\uparrow} c_{-\mathbf{k},yz,\uparrow} &= -\frac{1}{2}\left(a_{\mathbf{k},\uparrow} + b_{\mathbf{k},\uparrow}\right)\left(a_{-\mathbf{k},\uparrow} + b_{-\mathbf{k},\uparrow}\right) \\ &= -\frac{1}{2}\left(a_{\mathbf{k},\uparrow} a_{-\mathbf{k},\uparrow} + b_{\mathbf{k},\uparrow} b_{-\mathbf{k},\uparrow} + b_{\mathbf{k},\uparrow} a_{-\mathbf{k},\uparrow} + a_{\mathbf{k},\uparrow} b_{-\mathbf{k},\uparrow}\right). \end{aligned} \tag{29}$$

Adding $xz$ and $yz$ contributions:

$$\Delta_{\uparrow\uparrow}^{\text{intra}} \sim -(b_{\mathbf{k},\uparrow} a_{-\mathbf{k},\uparrow} + a_{\mathbf{k},\uparrow} b_{-\mathbf{k},\uparrow}). \tag{30}$$

Repeating the same steps for spin-down gives

$$\Delta_{\downarrow\downarrow}^{\text{intra}} \sim -(b_{\mathbf{k},\downarrow} a_{-\mathbf{k},\downarrow} + a_{\mathbf{k},\downarrow} b_{-\mathbf{k},\downarrow}).$$

Thus,

$$\Delta_{d_x}^{\text{intra}}(\mathbf{k}) \sim (b_{\mathbf{k},\uparrow} a_{-\mathbf{k},\uparrow} + a_{\mathbf{k},\uparrow} b_{-\mathbf{k},\uparrow}) - (b_{\mathbf{k},\downarrow} a_{-\mathbf{k},\downarrow} + a_{\mathbf{k},\downarrow} b_{-\mathbf{k},\downarrow})$$

This is a *pure interband* pairing in the SOC basis. Similarly the $d_y$ component resultss

$$\Delta_{d_y}^{\text{intra}}(\mathbf{k}) \sim \sum_{\ell=xz,yz} \left( c_{\mathbf{k},\ell,\uparrow} c_{-\mathbf{k},\ell,\uparrow} + c_{\mathbf{k},\ell,\downarrow} c_{-\mathbf{k},\ell,\downarrow} \right). \tag{31}$$

Repeating the same algebra as above yields

$$\Delta_{d_y}^{\text{intra}}(\mathbf{k}) \sim -(b_{\mathbf{k},\uparrow} a_{-\mathbf{k},\uparrow} + a_{\mathbf{k},\uparrow} b_{-\mathbf{k},\uparrow} + b_{\mathbf{k},\downarrow} a_{-\mathbf{k},\downarrow} + a_{\mathbf{k},\downarrow} b_{-\mathbf{k},\downarrow})$$

Again, this is purely interband. In summary:

$$\begin{aligned} \Delta_{d_z}^{\text{intra}}(\mathbf{k}) & \quad \text{pure \underline{intra}band,} \\ \Delta_{d_x}^{\text{intra}}(\mathbf{k}) & \quad \text{pure \underline{inter}band,} \\ \Delta_{d_y}^{\text{intra}}(\mathbf{k}) & \quad \text{pure \underline{inter}band.} \end{aligned} \tag{32}$$

Since pairing electrons in different energy bands costs energy due to spin-orbit coupling, we can conclude that the intra-orbital triplet pairing favors the out-of-plane component $d_z$, since it is the only one that remains *intraband* in the SOC basis. Instead, the in-plane components $d_x$ and $d_y$ are converted into *interband* pairing thus resulting energetically unfavorable.

### B. Inter-orbital pairing with odd symmetry under orbital exchange

The formal expressions of the three **d** vector components, when the superconducting pairing occurs between electrons of different orbitals in such a way to be antisymmetric under orbital exchange, is:

$$\Delta^{OO}_{d_x}(\mathbf{k}) = d_x^{even}(\mathbf{k})\big(c_{\mathbf{k},xz,\uparrow}c_{-\mathbf{k},yz,\uparrow} - c_{\mathbf{k},yz,\uparrow}c_{-\mathbf{k},xz,\uparrow}$$
$$-c_{\mathbf{k},xz,\downarrow}c_{-\mathbf{k},yz,\downarrow} + c_{\mathbf{k},yz,\downarrow}c_{-\mathbf{k},xz,\downarrow}\big), \quad (33)$$

$$\Delta^{OO}_{d_y}(\mathbf{k}) = d_y^{even}(\mathbf{k})\big(c_{\mathbf{k},xz,\uparrow}c_{-\mathbf{k},yz,\uparrow} - c_{\mathbf{k},yz,\uparrow}c_{-,xz,\uparrow}$$
$$+c_{\mathbf{k},xz,\downarrow}c_{-\mathbf{k},yz,\downarrow} - c_{\mathbf{k},yz,\downarrow}c_{-\mathbf{k},xz,\downarrow}\big), \quad (34)$$

$$\Delta^{OO}_{d_z}(\mathbf{k}) = d_z^{even}(\mathbf{k})\big(c_{\mathbf{k},xz,\uparrow}c_{-\mathbf{k},yz,\downarrow} - c_{\mathbf{k},yz,\uparrow}c_{-\mathbf{k},xz,\downarrow}$$
$$+c_{\mathbf{k},xz,\downarrow}c_{-\mathbf{k},yz,\uparrow} - c_{\mathbf{k},yz,\downarrow}c_{-\mathbf{k},xz,\uparrow}\big). \quad (35)$$

Due to orbital antysimmetry $d_x^{even}(\mathbf{k}), d_y^{even}(\mathbf{k}), d_z^{even}(\mathbf{k})$ are even functions of **k**, in such a way the overall pair wave function is antisymmetric under particle exchange.

Let us start considering the $d_z$ term, and let write it in the band basis. Using Eq. (22), we substitute:

$$\Delta^{OO}_{d_z}(\mathbf{k}) = \frac{i}{2}d_z^{even}(\mathbf{k})\big((b_{\mathbf{k},\uparrow} - a_{\mathbf{k},\uparrow})(b_{-\mathbf{k},\downarrow} + a_{-\mathbf{k},\downarrow})$$
$$-(b_{\mathbf{k},\uparrow} + a_{\mathbf{k},\uparrow})(a_{-\mathbf{k},\downarrow} - b_{-\mathbf{k},\downarrow})$$
$$+(a_{\mathbf{k},\downarrow} - b_{\mathbf{k},\downarrow})(a_{-\mathbf{k},\uparrow} + b_{-\mathbf{k},\uparrow})$$
$$-(a_{\mathbf{k},\downarrow} + b_{\mathbf{k},\downarrow})(b_{-\mathbf{k},\uparrow} - a_{-\mathbf{k},\uparrow})\big)$$

which finally gives

$$\Delta^{OO}_{d_z}(\mathbf{k}) = -id_z^{even}(\mathbf{k})\big(a_{\mathbf{k},\uparrow}a_{-\mathbf{k},\downarrow} - a_{\mathbf{k},\downarrow}a_{-\mathbf{k},\uparrow}$$
$$-b_{\mathbf{k},\uparrow}b_{-\mathbf{k},\downarrow} + b_{\mathbf{k},\downarrow}b_{-\mathbf{k},\uparrow}\big) \quad (36)$$

Thus, like for the intraorbital case, the $d_z$ component corresponds to a pure intraband coupling, which is energetically favored. Similarly,

$$\Delta^{OO}_{d_x}(\mathbf{k}) = id_x^{even}(\mathbf{k})\,\big(a_{\mathbf{k},\uparrow}b_{-\mathbf{k},\uparrow} - b_{\mathbf{k},\uparrow}a_{-\mathbf{k},\uparrow}$$
$$-b_{\mathbf{k},\downarrow}a_{-\mathbf{k},\downarrow} + a_{\mathbf{k},\downarrow}b_{-\mathbf{k},\downarrow}\big) \quad (37)$$

$$\Delta^{OO}_{d_y}(\mathbf{k}) = id_y^{even}(\mathbf{k})\,\big(a_{\mathbf{k},\uparrow}b_{-\mathbf{k},\uparrow} - b_{\mathbf{k},\uparrow}a_{-\mathbf{k},\uparrow}$$
$$+b_{\mathbf{k},\downarrow}a_{-\mathbf{k},\downarrow} - a_{\mathbf{k},\downarrow}b_{-\mathbf{k},\downarrow}\big) \quad (38)$$

In summary:

$$\Delta^{OO}_{d_z}(\mathbf{k}) \quad \text{pure \underline{intra}band,}$$
$$\Delta^{OO}_{d_x}(\mathbf{k}) \quad \text{pure \underline{inter}band,} \quad (39)$$
$$\Delta^{OO}_{d_y}(\mathbf{k}) \quad \text{pure \underline{inter}band.}$$

We can thus conclude that also in the case of interband pairing which is odd under orbital exchange, the **d** vector favors the alignement along the $z$ direction due to the atomic spin-orbit coupling.

### C. Inter-orbital pairing with even symmetry under orbital exchange

The formal expressions of the three **d** vector components the case of an interorbital spin triplet pairing with even symmetry under orbital exchange are:

$$\Delta^{OE}_{d_x}(\mathbf{k}) = d_x^{od}(\mathbf{k})\big(c_{\mathbf{k},xz,\uparrow}c_{-\mathbf{k},yz,\uparrow} + c_{\mathbf{k},yz,\uparrow}c_{-\mathbf{k},xz,\uparrow}$$
$$-c_{\mathbf{k},xz,\downarrow}c_{-\mathbf{k},yz,\downarrow} - c_{\mathbf{k},yz,\downarrow}c_{-\mathbf{k},xz,\downarrow}\big), \quad (40)$$

$$\Delta^{OE}_{d_y}(\mathbf{k}) = d_y^{odd}(\mathbf{k})\big(c_{\mathbf{k},xz,\uparrow}c_{-\mathbf{k},yz,\uparrow} + c_{\mathbf{k},yz,\uparrow}c_{-,xz,\uparrow}$$
$$+c_{\mathbf{k},xz,\downarrow}c_{-\mathbf{k},yz,\downarrow} + c_{\mathbf{k},yz,\downarrow}c_{-\mathbf{k},xz,\downarrow}\big), \quad (41)$$

$$\Delta^{OE}_{d_z}(\mathbf{k}) = d_z^{odd}(\mathbf{k})\big(c_{\mathbf{k},xz,\uparrow}c_{-\mathbf{k},yz,\downarrow} + c_{\mathbf{k},yz,\uparrow}c_{-\mathbf{k},xz,\downarrow}$$
$$+c_{\mathbf{k},xz,\downarrow}c_{-\mathbf{k},yz,\uparrow} + c_{\mathbf{k},yz,\downarrow}c_{-\mathbf{k},xz,\uparrow}\big). \quad (42)$$

where $d_x^{odd}(\mathbf{k}), d_y^{odd}(\mathbf{k}), d_z^{odd}(\mathbf{k})$, this time, are odd functions of **k**, due to the overall antisymmetry of the pair function. Using Eq. (22), we get

$$\Delta^{OE}_{d_x}(\mathbf{k}) = -id_x^{odd}(\mathbf{k})\big(a_{\mathbf{k},\uparrow}a_{-\mathbf{k},\uparrow} - b_{\mathbf{k},\uparrow}b_{-\mathbf{k},\uparrow}$$
$$-a_{\mathbf{k},\downarrow}a_{-\mathbf{k},\downarrow} + b_{\mathbf{k},\downarrow}b_{-\mathbf{k},\downarrow}\big) \quad (43)$$

$$\Delta^{OE}_{d_y}(\mathbf{k}) = id_y^{odd}(\mathbf{k})\big(a_{\mathbf{k},\uparrow}a_{-\mathbf{k},\uparrow} - b_{\mathbf{k},\uparrow}b_{-\mathbf{k},\uparrow}$$
$$+a_{\mathbf{k},\downarrow}a_{-\mathbf{k},\downarrow} - b_{\mathbf{k},\downarrow}b_{-\mathbf{k},\downarrow}\big) \quad (44)$$

$$\Delta^{OE}_{d_z}(\mathbf{k}) = id_z^{odd}(\mathbf{k})\big(b_{\mathbf{k},\uparrow}a_{-\mathbf{k},\downarrow} - a_{\mathbf{k},\uparrow}b_{-\mathbf{k},\downarrow}$$
$$-b_{\mathbf{k},\downarrow}a_{-\mathbf{k},\uparrow} + a_{\mathbf{k},\downarrow}b_{-\mathbf{k},\uparrow}\big) \quad (45)$$

Therefore:

$$\Delta^{OE}_{d_x}(\mathbf{k}) \quad \text{pure \underline{intra}band,}$$
$$\Delta^{OE}_{d_y}(\mathbf{k}) \quad \text{pure \underline{intra}band,} \quad (46)$$
$$\Delta^{OE}_{d_z}(\mathbf{k}), \quad \text{pure \underline{inter}band.}$$

Thus, in the case of interband pairing which is even under orbital exchange, the planar components $d_x, d_y$ remain purely intraband, while in contrast, $d_z$ is energetically suppressed by the SOC, yielding *a planar multicomponent triplet order parameter*. This is the most relevant state for LiFeAs, since it can be responsible for the occurrence of non trivial vortex textures.

### D. Nematic triplet state

We classify the possible nematic states for LiFeAs that are compatible with the orbital symmetric interorbital order parameter under the full tetragonal group $D_{4h}$. In the context of tetragonal $D_{4h}$ symmetry, a superconducting state is defined as *nematic* if the order parameter spontaneously breaks the discrete four-fold rotational symmetry ($C_4$) of the lattice down to a two-fold axis $C_2$. In other words, a nematic state selects a preferred in-plane direction, resulting in an anisotropic superconducting gap in the $xy$ plane. This symmetry reduction typically manifests when the system populates a multi-dimensional irreducible representation, such as the two-dimensional $E_u$ representation, and selects a non-trivial linear combination of its basis functions. For the interorbital symmetric case where the **d**-vector components lie in the $xy$-plane, the basis for $E_u$ can be spanned by:

$$\mathbf{d}_1(\mathbf{k}) = (f_1(\mathbf{k}), 0, 0), \quad \mathbf{d}_2(\mathbf{k}) = (0, f_2(\mathbf{k}), 0). \tag{47}$$

where $f_{1,2}(\mathbf{k})$ are odd functions of $\mathbf{k}$. The general form of the order parameter is then a linear combination $\mathbf{d}(\mathbf{k}) = \eta_1 \mathbf{d}_1 + \eta_2 \mathbf{d}_2$. Although any such combination belongs to the $E_u$ manifold, the specific choice of the coefficients $(\eta_1, \eta_2)$ determines the symmetry of the resulting superconducting phase.

We consider two primary candidates for the **d**-vector. The first is defined as $\mathbf{d}(\mathbf{k}) = \Delta_0(\sin k_z, \sin k_z, 0)$, which identifies the [110] diagonal as the nematic director. This state remains fully gapped on the experimentally relevant Fermi surfaces away from $k_z = 0$.

The second candidate arises from the in-plane basis functions $\mathbf{d}_1 = (\sin k_x \cos k_y, 0, 0)$ and $\mathbf{d}_2 = (0, \sin k_y \cos k_x, 0)$. While in this second case, a fully symmetric combination $\mathbf{d}_1 + \mathbf{d}_2$ would preserve the tetragonal invariance, the spontaneous selection of a single component, or an unequal weighting of the two, signals the onset of a nematic phase. Due to the combination of the two components, this state can be fully gapped on the Fermi surfaces, consistent with the experimentally observed saturated behaviour of the depolarization rate.

## V. TEMPERATURE EVOLUTION OF THE SKIRMIONIC VORTEX LATTICE

### A. Deriving the Ginzburg-Landau model

In the paper we use a transformed version of the Gibbs free energy used in [25, 26]. To derive this, we start from the 3-dimensional anisotropic Ginzburg-Landau free energy density used in [27] and derived microscopically in [28],

$$\mathscr{F} = \sum_{s=\pm} \left\{ -|\Delta_s|^2 + |D_x \Delta_s|^2 + |D_y \Delta_s|^2 + \beta_z |D_z \Delta_s|^2 \right.$$
$$\left. + \beta_\perp \overline{D_{-s} \Delta_s} D_s \Delta_{-s} + \frac{1}{2}|\Delta_s|^4 + \frac{\gamma}{2}|\Delta_s|^2|\Delta_{-s}|^2 \right\}, \tag{48}$$

where $\Delta_\pm = |\Delta_\pm| e^{i\varphi_\pm}$ are the two complex order parameters and we have covariant derivatives $D_\pm = D_x \pm i D_y$, where the spatial covariant derivatives are $D_i = -(i/\eta)\partial_i + a_i$, and $a$ is the gauge field with associated magnetic field $b = \nabla \times a$.

It will be useful for applying our algorithm to transform the free energy to the general form presented in [29]. We can rescale the theory presented in (48) using the following rescaled fields,

$$\mathfrak{F} = \frac{1}{2}\eta^2 \mathscr{F}, \tag{49}$$

$$\mathbf{A} = -\frac{1}{\eta} a, \tag{50}$$

$$\psi_1 = \Delta_+, \tag{51}$$

$$\psi_2 = \Delta_-. \tag{52}$$

If we then write down the resulting free energy $F = \int_{\mathbb{R}^3} \mathfrak{F}$ in terms of these transformed fields, it is in the desired general form used throughout the paper,

$$F = \int_{\mathbb{R}^3} \left\{ \frac{1}{2} Q_{ij}^{\alpha\beta} \overline{D_i \psi_\alpha} D_j \psi_\beta + \frac{1}{2}|\mathbf{B}|^2 + F_p(\rho_\alpha, \theta_{12}) \right\} d^3x, \tag{53}$$

where $D_i = \partial_i - iA_i$ is the transformed covariant derivative associated with the new gauge field $A_i$, leading to the local magnetic field $\mathbf{B} = \nabla \times \mathbf{A}$. The two complex fields are now $\psi_\alpha = \rho_\alpha e^{i\varphi_\alpha}$. Note that Greek indices $\alpha = 1, 2$ will always enumerate components of the transformed order parameter and Latin indices $i = 1, 2, 3$ indicate spatial components, while summation over repeated indices is implied for both.

The gradient terms are represented by the constant matrices $Q^{\alpha\beta}$,

$$Q^{11} = Q^{22} = \begin{pmatrix} 1 & 0 & 0 \\ 0 & 1 & 0 \\ 0 & 0 & \beta_z \end{pmatrix}, \quad Q^{12} = \beta_\perp \begin{pmatrix} 1 & i & 0 \\ i & -1 & 0 \\ 0 & 0 & 0 \end{pmatrix}, \tag{54}$$

where $\beta_\perp$ and $\beta_z$ are the same positive parameters as in (48) which must be fixed. Note that $\beta_\perp$ will only appear in the equations of motion if vortices are not translation invariant in the $c$-direction. $F_p$ collects together the transformed potential terms, which due to gauge invariance, can depend only on the condensate magnitudes $\rho_\alpha$ and the phase difference between the condensates $\theta_{12} := \theta_1 - \theta_2$,

$$F_p = \frac{\eta^2}{2} \left( -|\psi_1|^2 - |\psi_2|^2 + \frac{1}{2}(|\psi_1|^4 + |\psi_2|^4) + \gamma |\psi_1|^2 |\psi_2|^2 \right). \tag{55}$$

We note that the homogeneous ground state (minimal energy degree 0) solution is given by the values of $\rho_\alpha$ that minimize $F_p$. We assume that $\gamma < 1$ such that the model has a $U(1) \times U(1)$ symmetry with a single ground state,

$$u_1 = u_2 = \frac{1}{\sqrt{1+\gamma}}. \tag{56}$$

It is worth noting that while the model does not have rotation symmetry, it does have a mixed symmetry, resulting from breaking the $U(1) \times U(1)$ symmetry of $F_p$ if $\beta_\perp \neq 0$. In particular $F$ has the symmetry $F(\psi, A) = F(\tilde{\psi}, \tilde{A})$ where,

$$\tilde{\psi}(x) = S\psi(R^{-1}x), \quad \tilde{A}(x) = RA(R^{-1}x) \tag{57}$$

and

$$R = \begin{pmatrix} \cos\vartheta & -\sin\vartheta & 0 \\ \sin\vartheta & \cos\vartheta & 0 \\ 0 & 0 & 1 \end{pmatrix}, \quad S = \begin{pmatrix} e^{i\vartheta} & 0 \\ 0 & e^{-i\vartheta} \end{pmatrix}. \tag{58}$$

This couples spatial rotations about the $c$ axis with rotations of the phase difference, meaning any solutions we find can be rotated freely (we assume any extra terms introducing rotational symmetry breaking are weak).

Individual vortices, in an infinite superconductor, are local minima of $F$, however they appear in experiment as periodic lattice structures. We simulate such structures as local minimisers of the Gibbs free energy, which we define,

$$G = F - \boldsymbol{H} \cdot \boldsymbol{B} + \boldsymbol{H}^2, \tag{59}$$

where $\boldsymbol{H}$ is the applied magnetic field to the system (we will always assume in the $c$-direction). In the next section it is shown that for fields applied in the $c$-direction we can reduce to a 2-dimensional periodic lattice in the basal $a,b$-plane and consider only excitations of the local magnetic field $\boldsymbol{B}$ in the $c$-direction also [30]. The other key thermodynamic quantity is the temperature of the system $T/T_c$ (which we write relative to the critical temperature $T_c$), which appears in the potential in (55), through the parameter,

$$\alpha(T) = (1 - T/T_c)\alpha_0. \tag{60}$$

We can account for this, after a rescaling, by using the generalized potential term,

$$F_p = \frac{\eta^2}{2}\left(-(1-T/T_c)(|\psi_1|^2 + |\psi_2|^2) + \frac{1}{2}(|\psi_1|^4 + |\psi_2|^4)\right.$$
$$\left. + \gamma|\psi_1|^2|\psi_2|^2\right), \tag{61}$$

which now depends on $T/T_c$. We will assume throughout that $T < T_c$ such that the system is superconducting. It is informative to make the field transformation $\rho_\alpha^2 \mapsto (1-T/T_c)\rho_\alpha^2$, followed by a spatial rescaling $x \mapsto x/\sqrt{1-T/T_c}$ and an energy rescaling $G \mapsto G/(1-T/T_c)$. This leaves the rescaled Gibbs free energy, dependent on temperature $T$ and applied field $\boldsymbol{H}$,

$$G = \int_\Omega \left\{ \frac{1}{2} Q_{ij}^{\alpha\beta}\overline{D_i\psi_\alpha}D_j\psi_\beta + \frac{1}{2}\left|B - \frac{H}{1-T/T_c}\right|^2 + F_p(\psi_\alpha)\right\}. \tag{62}$$

Note we are now integrating over the unit cell of the given lattice $\Omega$. This reformulation makes it clear that a change in temperature (when $H < H_{c_2}$ and $T < T_c$) is qualitatively indistinguishable from a change in the applied field $H$ (up to spatial and field re-scalings). The resulting equations of motion are then obtained by varying $G$ w.r.t. $(\psi, A)$,

$$Q_{ij}^{\alpha\beta}D_iD_j\psi_\beta = 2\frac{\partial F_p}{\partial \overline{\psi}_\alpha}, \tag{63}$$

$$\partial_i(\partial_j A_i - \partial_i A_j) = J_i, \tag{64}$$

where the total supercurrent is defined as,

$$J_i := Im(Q_{ij}^{\alpha\beta}\overline{\psi}_\alpha D_j\psi_\beta). \tag{65}$$

Note that these are completely independent of $\boldsymbol{H}$ and $T$. This is not a surprise as they are thermodynamic quantities that will have an effect in finding the optimal unit cell geometry $\Omega$ (see the lattice section).

### B. Skyrmions in Ginzburg-Landau

Skyrmions form when integer flux vortices split into spatially separated fractional (often called half-quanta) vortices in each component $\psi_\alpha$, such that their zeros ($\psi_\alpha = 0$) do not coincide. Since any such field configuration never attains the value $(\psi_1, \psi_2) = (0, 0) \in \mathbb{C}^2$, we may construct from it a gauge invariant field $\Phi : \mathbb{R}^2 \to S^2$,

$$\Phi(x_1, x_2) = \frac{(\bar{\psi}_1\psi_2 + \psi_1\bar{\psi}_2, i(\bar{\psi}_1\psi_2 - \psi_1\bar{\psi}_2), |\psi_2|^2 - |\psi_1|^2)}{|\psi_1|^2 + |\psi_2|^2}, \tag{66}$$

which is referred to as the Skyrme field. This field takes values on a sphere (directions in 3-dimenions), similar to the magnetization vector field of a magnetic skyrmion. It can be shown (using Stoke's theorem) that when $\Phi \neq 0$ that the local magnetic field $\boldsymbol{B}$, which is topologically preserved, is equivalent to the topological degree (or winding number) of $\Phi$, which counts the skyrmions.

The individual phases $\varphi_\alpha$ wind around fractional vortices in the two components $\psi_\alpha = \rho_\alpha e^{i\varphi_\alpha}$, meaning when separated into a skyrmion the phase difference $\theta_{12} = \varphi_1 - \varphi_2$ must wind by $2\pi$ between them. This is simply the director of the nematic system winding. This is energetically penalized by the standard (non-coupled) gradient terms of the Ginzburg-Landau, hence Skyrmions rarely form. However, it has been shown that particular coupling terms in the potential [31] or gradient [32] can favor splitting. For our model the splitting is driven by two parameters, $\gamma$ in the potential (55) and $\beta_\perp$ in the gradient coupling matrices (54). The potential term contributes less energy after splitting as this minimises the overlap of the condensate magnitudes which is squared. For the $\beta_\perp$ term, consider a co-centered fractional vortices such that the term integrates to zero. By splitting the fractional vortices this term can become negative.

### C. Fixing Ginzburg-Landau parameters

While we have derived the model that we will simulate we don't yet know what parameters are sensible for LiAsFe. As Ginzburg-Landau theory is just an effective theory we cannot make precise quantitative predictions, but we can make qualitative predictions and compare them to the experimental results. We will fix the two parameters of the model by ensuring that the model approximately gives the correct critical field values and length scales.

To find $H_{c_2}$ for an anisotropic superconductor there are multiple methods that we will not present here. However finding $H_{c_1}$ is much harder. For anisotropic superconductors this was recently solved in [30] and we follow the method presented there.

The asymptotic length scales for any model of the general form (53) can be found using the method described in [33]. While this gives the length scales, when $\beta_\perp \neq 0$ the matter and magnetic length scales will hybridize and be direction dependent in general [25]. This makes comparing the penetration depth and coherence length difficult. To make progress we will make use of the London model which in general for a model of the form of (53) is,

$$F_{lon} = \frac{1}{2}\mathcal{Q}_{ij}A_iA_j + \frac{1}{2}\left(\partial_x A_y - \partial_y A_x\right)^2, \tag{67}$$

where we have used $\mathcal{Q}_{ij} = u_\alpha u_\beta Q_{ij}^{\alpha\beta}$ and we know that $\mathcal{Q}$ is symmetric and positive definite. Hence, we now write this in terms of a new orthonormal basis given by the eigenvectors $\xi_i$ of $\mathcal{Q}$ such that,

$$(A_x, A_y)^T = A_1 \xi_1 + A_2 \xi_2, \tag{68}$$

which leads to,

$$F_{lon} = \frac{1}{2}\mu_i^2 A_i^2 + \frac{1}{2}(\partial_1 A_2 - \partial_2 A_1)^2, \tag{69}$$

where $\mu_i^2$ are the eigenvalues of $\mathcal{Q}$ which must be positive and real. Hence, the penetration depth is anisotropic and is $1/\mu_i$ in the $\xi_i$ direction.

To find the approximate coherence lengths (of which there are two due to the two components), we assume that the gauge field $A$ is constant. If we then consider small values $\epsilon_\alpha = |\psi_\alpha| - u_\alpha$,

$$F_{lin} = \frac{1}{2}Q_{ij}^{\alpha\beta}\partial_i\epsilon_\alpha\partial_j\epsilon_\beta + \frac{1}{2}\epsilon_\alpha \mathcal{H}_{\alpha\beta}\epsilon_\beta, \tag{70}$$

where $\mathcal{H}$ is the Hessian of the potential term $F_p$,

$$\mathcal{H}_{\alpha\beta} := \left.\frac{\partial^2 F_p}{\partial|\psi_\alpha|\partial|\psi_\beta|}\right|_{(u_1, u_2, \delta_{12})} \tag{71}$$

and $\delta_{12}$ is the vacuum phase difference value. We first start by picking a direction in the plane written $n = (\cos\theta, \sin\theta)^T$. Then we can assume that orthogonal changes are negligable to give,

$$F_{lin} = \frac{1}{2}\mathfrak{Q}_{\alpha\beta}\partial_n\epsilon_\alpha\partial_n\epsilon_\beta + \frac{1}{2}\epsilon_\alpha \mathcal{H}_{\alpha\beta}\epsilon_\beta, \tag{72}$$

where $\mathfrak{Q}_{\alpha\beta} = n_i n_j Q_{ij}^{\alpha\beta}$, which gives the linear ODE to solve,

$$\partial_n^2 \epsilon_\alpha + (\mathfrak{Q}^{-1}\mathcal{H})_{\alpha\beta}\epsilon_\beta = 0. \tag{73}$$

Hence, we can now extract the two condensate length scales as $1/\nu_\alpha$, where $\nu_\alpha^2$ are positive real eigenvalues of the matrix $\mathfrak{Q}^{-1}\mathcal{H}$.

For the material we are interested in we make use of the following data,

- $\lambda_c/\xi_c \approx 50$,
- $H_{c_2}/H_{c_1} \approx 700$.

If we assume a radially symmetric configuration then $Q^{12} = 0$, hence for $H \parallel \hat{c}$ we have,

$$Q^{11} = Q^{22} = \begin{pmatrix} 1 & 0 \\ 0 & 1 \end{pmatrix}, \tag{74}$$

such that we can calculate the length scales directly to be,

$$\lambda_c = \frac{1}{\sqrt{2}u} = \sqrt{\frac{1+\gamma}{2}}. \tag{75}$$

For the coherence length we have,

$$\mathcal{H} = \eta^2 \begin{pmatrix} -1 + u^2(3+\gamma) & 2\gamma u^2 \\ 2\gamma u^2 & -1 + u^2(3+\gamma) \end{pmatrix}, \tag{76}$$

$$= \frac{2\eta^2}{1+\gamma}\begin{pmatrix} 1 & \gamma \\ \gamma & 1 \end{pmatrix} \tag{77}$$

where the eigenvectors (assuming $\gamma \neq 0$) are of the form,

$$\nu_+ = \begin{pmatrix} 1 \\ 1 \end{pmatrix} \qquad \nu_- = \begin{pmatrix} -1 \\ 1 \end{pmatrix}, \tag{78}$$

and the corresponding eigenvalues are $\mu_\pm^2 = \frac{2\eta^2}{1+\gamma}(1\pm\gamma)$. Hence the coherence lengths are then given as,

$$\xi_\pm = \frac{1}{\mu_\pm} = \sqrt{\frac{1+\gamma}{2\eta^2(1\pm\gamma)}}, \tag{79}$$

so if we are interested in the ratio $\lambda_c/\xi_c$ and we want this as large as possible, then take $\xi_c = \xi_+ = 1/\sqrt{2}\eta$ and then,

$$\frac{\lambda_c}{\xi_c} = \eta\sqrt{1+\gamma}, \tag{80}$$

If we take $\gamma \sim 1$ then we end up with $\eta \approx 50/\sqrt{2} \approx 35$.

We know that $H_{c_2} \approx 1.2676\eta^2$ from [25]. On the other hand calculating $H_{c_1}$ is inherently a non-linear problem as,

$$H_{c_1} = \min_N \frac{F_N}{2\pi N}, \tag{81}$$

where $F_N$ is the free enery of the $N$-vortex global minimiser and hence inherently non-linear. As we are working with a qualitative model, we only seek to fit approximate quantities and hence we assume that,

$$H_{c_1} \approx \frac{F_1}{2\pi} \approx \frac{F_1^{lon}}{2\pi}, \tag{82}$$

where $F_1$ is the free energy of a single vortex (skyrmion), which we approximate by assuming that $\beta_\perp = 0$ and taking the London model. Approximating the free energy of a vortex from the London model is a well studied problem,

$$F_1^{lon} = \frac{\pi}{\lambda^2}\ln\left(\frac{\lambda}{\xi}\right), \tag{83}$$

which leads to the approximation $H_{c_1} \approx \frac{1}{1+\gamma}ln(\eta\sqrt{1+\gamma})$.

Note that we have used a lot of approximations to approximate the length scales and critical field values above. This was to approximate the parameters only. In reality the length scales are far more complicated and highly coupled. This coupling is required to explain many of the features of unconventional vortex properties in this and other models [25, 29, 34].

The Ginzburg-Landau model is an effective model, that approximates the behavior of vortices in superconductors. We do not expect this model to be quantitatively accurate for the experiment we consider in this paper. We therefore focus purely on the qualitative results in the form of vortex lattice structure, and the phases this goes through. As we are only interested in qualitative results we do not need to be precise with the parameters of the model and approximate values are sufficient.

### D. Lattice solutions of the Ginzburg-Landau model

To compare with experiment, we seek lattice (periodic) solutions that minimise the Gibbs free energy per unit volume. As $\mu$SR experiments probe the bulk of the superconductor [35] we want to find the bulk solution with no boundary effects. The standard approach is to take a large (costly to simulate) box as a system with a number of vortices and assume the box is big enough that the boundary effects are negligible. This is not a good approach for $\mu$SR as the boundaries affect the signal we want to compare with experiment. In addition, as we seek an unconventional vortex lattice, we don't know it's symmetry ahead of time, unlike for an Abrikosov lattice. Hence, we seek the simplest unit cell that is tessellated to form the lattice solution.

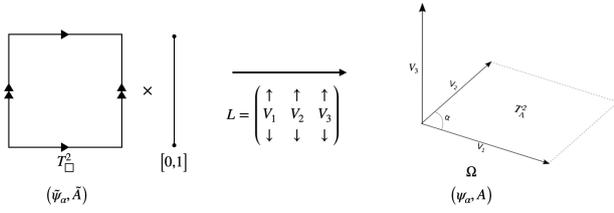

Fig. SI-7. The matrix $L$ maps between the fields used in simulation $(\tilde{\psi}_\alpha, \tilde{A})$ defined on a square torus of unit area $T_\square^2$ and the physical fields $(\psi_\alpha, A)$ defined over the physical lattice unit cell $T_\Lambda^2$.

We will follow the approach of [30], though a less general approach was first presented in [25]. In particular, we seek the configuration $(\Omega, \psi, A)$ that is the global minimizer of $G[\Omega, \psi, A]/|\Omega|$, where $\Omega$ is the unit cell (see Figure SI-7) and hence $|\Omega|$ is the volume of the unit cell.

We first choose an oriented basis $[v_1, v_2, v_3]$ for $\mathbb{R}^3$, giving the coordinates $X_i$ such that,

$$x = LX = X_1 v_1 + X_2 v_2 + X_3 v_3, \tag{84}$$

where $L$ is the matrix whose columns are the chosen basis. Can then explicitly define the unit cell $\Omega$ as the parallelepiped spanned by $[v_1, v_2, v_3]$, with cell coordinates $X_i$ and volume $|\Omega| = \det L$.

We will impose translation symmetry in the direction $v_3$ and, without loss of generality, assume that $v_1 \cdot v_3 = v_2 \cdot v_3 = 0$ such that,

$$v_3 = \frac{v_1 \times v_2}{|v_1 \times v_2|^2} \tag{85}$$

so that the frame is automatically positively oriented, and the unit cell has volume 1.

In order to allow non-zero magnetic flux, the fields are periodic only up to gauge with the following boundary conditions,

$$\psi_\alpha(X_1 + 1, X_2) = \psi_\alpha(X_1, X_2) e^{i2\pi N X_2}, \tag{86}$$
$$\psi_\alpha(X_1, X_2 + 1) = \psi_\alpha(X_1, X_2), \tag{87}$$
$$A_1(X_1 + 1, X_2) = A_1(X_1, X_2), \tag{88}$$
$$A_2(X_1 + 1, X_2) = A_2(X_1, X_2) + 2\pi N, \tag{89}$$
$$A_3(X_1 + 1, X_2) = A_3(X_1, X_2), \tag{90}$$
$$A_i(X_1, X_2 + 1) = A_i(X_1, X_2). \tag{91}$$

Hence we have applied even winding purely along the $X_2$ boundary of the torus $T_\square^2$ (fixing some of the gauge freedom of the e.o.m.). This choice leaves all gauge invariant quantities $\rho_\alpha$, $\varphi_{\alpha\beta}$, $J_i$ and $B$ doubly periodic as required. Note that while not immediately obvious (as the basis is not orthonormal) the magnetic field is now $B = B_1 v_1 + B_2 v_2 + B_3 v_3$.

Finally, we write the Gibbs free energy in the new coordinate system,

$$G = \int_\Omega \left\{ \frac{1}{2} M_{ki} Q_{ij}^{\alpha\beta} M_{jl}^T \overline{D_{X_k} \psi_\alpha} D_{X_l} \psi_\beta + \frac{1}{2} \left| B - \frac{H}{1 - T/T_c} \right|^2 + F_P(\rho, \vartheta_1 \tag{92}$$

where $M = L^{-1}$ and for convenience we will denote,

$$H^*(T) = \frac{H}{1 - T/T_c}. \tag{93}$$

We can then simplify the above expression using some of the assumptions we have made [30].

$$\langle G \rangle = \frac{G}{|\Omega|} = \frac{1}{2} M_{ki} P_{ki,lj} M_{lj} + \frac{1}{2} Tr(L \mathbb{B} L^T) - 2N\pi H_i^* L_{i3}$$
$$+ \frac{1}{2} |H^*|^2 + \int_{[0,1]^2} F_p(\psi), \tag{94}$$

where we have introduced,

$$P_{ki,lj} = Re \int_{[0,1]^2} Q_{ij}^{\alpha\beta} \overline{D_{X_k} \psi_\alpha} D_{X_l} \psi_\beta \, dX_1 \, dX_2, \tag{95}$$

$$\mathbb{B}_{ij} = \int_{[0,1]^2} B_i B_j \, dX_1 \, dX_2. \tag{96}$$

Using the above formulation for $\langle G \rangle$ our numerical scheme will alternate between minimising $G[\psi, A]$ w.r.t. the fields $(\psi, A)$ while keeping the unit cell $L$ fixed and then fixing the field configuration and minimising $\langle G \rangle$ w.r.t. $L \in SL(3, \mathbb{R})$. The first part we will use the arrested Newton flow method, which is described in detail in [29]. For the second part we consider the first 3 terms of (94) as these are the only ones dependent on $L$ where,

$$L = \begin{pmatrix} v_1 & v_2 & \dfrac{v_1 \times v_2}{|v_1 \times v_2|^2} \end{pmatrix}. \tag{97}$$

We assume that the plane $T_\square^2$ in figure SI-7 spanned by $v_1, v_2$ is orthogonal to $v_3$ and that $\Omega$ has unit volume, which is equivalent to the conditions,

$$\det L = 1, \tag{98}$$
$$L_{i1} L_{i3} = 0, \tag{99}$$
$$L_{i2} L_{i3} = 0. \tag{100}$$

As the condition in (98) is cubic, we must minimize this numerically (had it been quadratic as in [25, 26] we could have done this step explicitly).

The numerical goal is now, given a fixed configuration $(\psi_\alpha, A)$, to minimize $\langle G \rangle$ in (94) over the space of matricies $\mathscr{C} \subset GL(3, \mathbb{R})$ that obey the constraints (98)-(100). Let $L(t)$ be a curve in $\mathscr{C}$ where,

$$M(t) L(t) = I_3, \tag{101}$$
$$\dot{M}(0) L(0) + M(0) \dot{L}(0) = 0, \tag{102}$$
$$\dot{M}(0) = -M(0) \dot{L}(0) M(0), \tag{103}$$

leading to,

$$\frac{d}{dt}\bigg|_{t=0} \langle G \rangle (L(t)) = \epsilon_{ik} \left( -M_{qp} P_{qp,lj} M_{li} M_{kj} + L_{ij} \mathbb{B}_{jk} - 2N\pi H_i^* \delta_{k3} \right). \tag{104}$$

Hence, the gradient of $\langle G \rangle : \mathscr{C} \to \mathbb{R}$ at $L \in \mathscr{C}$ tangent to $\mathscr{C}$ is,

$$(\text{grad} \langle G \rangle)_{ik} = P_{\mathscr{C}} \left( -M_{qp} P_{qp,lj} M_{li} M_{kj} + L_{ij} \mathbb{B}_{jk} - 2N\pi H_i^* \delta_{k3} \right), \quad (105)$$

where $P_{\mathscr{C}}$ denotes orthogonal projection tangent to $\mathscr{C} \subset \mathbb{R}^9$. The projector $P_{\mathscr{C}} : \mathbb{R}^9 \to T_L \mathscr{C}$ is straightforward to construct numerically via a Gramm-Schmidt algorithm, starting from the basis of coordinate basis vectors for $T_L \mathscr{C}$ defined by the parametrization (97).

Once we have the projector we can minimize $\langle G \rangle$ to a given tolerance $g_{tol}$ with respect to the unit cell $L$ (for a fixed field configuration $(\psi_\alpha, A)$) via a simple gradient descent algorithm. Hence, we evolve $L$ as,

$$L \mapsto L - dt \, (\text{grad} \langle G \rangle) \quad (106)$$

repeating until the absolute values of all components of grad $\langle G \rangle$ are smaller than $g_{tol}$.

We must also find the fields $(\psi, A)$ that minimise $\langle G \rangle$ subject to a fixed $L$ (or equivalently minimisers of $G$). Note that any standard numerical method for field theories would work, we made use of the arrested Newton flow algorithm. This approach (described in detail in [36]) solves for the motion of a particle in the configuration space under the potential $\langle G \rangle$. Then if the direction of the force on the particle opposes its velocity $(\dot{\psi}, \dot{A}) \cdot \text{grad}_{(\psi, A)} \langle G \rangle > 0$, then we set $(\dot{\psi}, \dot{A}) = 0$ and restart the flow.

### E. Extracting the probability density

To compare the lattice (periodic) minimisers of (94) with $\mu$SR experiments, we must convert the unit cell $\Omega$ and numerical fields $(\psi, A)$ to a comparable signal, namely the probability density distribution $p(B)$. The approach we outlined above is ideal for this, as it produces the field configuration for a single unit cell. We use a standard kernel density estimation [35, 37] to approximate a continuous distribution from our discrete field configuration. Essentially, we must map the discrete magnetic field configuration on a unit cell $\Omega$ to a continuous probability density function $p(B)$ describing the distribution of local magnetic field values.

Let $B_{ij}{}_{i,j=1}^n$ denote the magnetic field on a uniform grid of $n^2$ points in $\Omega$. This induces an empirical measure [38],

$$\mu_n(B) = \frac{1}{n^2} \sum_{i,j=1}^n \delta(B - B_{ij}), \quad (107)$$

which is a discrete probability measure supported on the sampled field values. Since $\mu_n$ is singular, we approximate it by a smooth density using a kernel density estimation [35, 37], i.e. by convolution with a Gaussian kernel of bandwidth $h > 0$. This results in an approximation for the probability density,

$$p(B) = \frac{1}{hn^2} \sum_{i=1}^n \sum_{j=1}^n \frac{1}{\sqrt{2\pi}} e^{-\frac{1}{2}\left(\frac{B - B_{ij}}{h}\right)^2}. \quad (108)$$

Hence, $p(B)$ is a smooth approximation to the distribution of magnetic field values over $\Omega$, satisfying $\int_{\mathbb{R}} p(B) \, dB = 1$.

The bandwidth $h$ controls the smoothing, hence as $h \to 0$, $p(B)$ converges to $\mu_n(B)$, while for finite $h$ it provides a regularized approximation whose resolution should be chosen relative to the grid spacing and sample size $n^2$.

---

* Corresponding author: gianrico.lamura@spin.cnr.it
† Corresponding author: twinyard001@dundee.ac.uk
‡ Corresponding author: paola.gentile@spin.cnr.it
§ Corresponding author: tshiroka@phys.ethz.ch